\RequirePackage{amsmath}
\documentclass[6pt]{article}
\usepackage[a4paper, total={6in, 8in}]{geometry}
\usepackage{booktabs} 
\usepackage{fullpage}
\usepackage{graphicx}
\usepackage{pgfplots}
\usepackage[utf8]{inputenc}
\usepackage{tikz}
\usepackage{dblfloatfix}
\usepackage{float}
\usepackage{textgreek}
\usepackage{multicol}
\usepackage{algorithm}
\usepackage{relsize}
\usepackage{mathtools, bm, amsfonts, amssymb, lmodern}

\usepackage{hyperref}
\usepackage{subfig}
\usepackage{verbatim}
\usepackage{algorithmic}
\usepackage[inline]{enumitem} 
\usepackage{setspace}

\definecolor{blue}{HTML}{1F77B4}
\definecolor{orange}{HTML}{FF7F0E}
\definecolor{green}{HTML}{2CA02C}

\pgfplotsset{compat=1.14}

\newcommand\MEETtitle[1]{\Large \bf \hskip2.25pc \parbox{.8\textwidth}{ \noindent%
		\large \bf \begin{center} #1 \end{center}\rm } \vskip.1in \rm\normalsize }

\newcommand\MEETauthor[1]{\hskip2.25pc \parbox{.8\textwidth}{ \noindent%
		\normalsize \bf \begin{center} #1 \end{center}\rm } \vskip-1pc }

\let\title\MEETtitle
\let\author\MEETauthor

\let\address\MEETaddress
\let\email\MEETemail

\begin{document}	
	\title{Zero-Inflated Poisson Cluster-Weighted Models: Properties and Applications}

	\author{Kehinde Olobatuyi}
	\address{Department of Statistics and Mathematics Finance, \\ University of Milan-Bicocca, Milan, Italy.}
	\email{k.olobatuyi@campus.unimib.it}
	
\begin{abstract}
	\noindent In this paper, I propose a new class of Zero-Inflated Poisson models into the family of Cluster Weighted Models (CWMs) called Zero-Inflated Poisson CWMs (ZIPCWM). ZIPCWM extends Poisson cluster weighted models and other mixture models. I propose an Expectation-Maximization (EM) algorithm via an iteratively reweighted least squares for the model. I theoretically and analytically investigate the identifiability of the proposed model through an extensive simulation study. Parameter recovery, classification assessment, and performance of different information criteria are investigated through broad simulation design. ZIPCWM is applied to real data which accounts for excess zeros of over $40\%$. We explore the classification performance of ZIPCWM, Fixed Zero-inflated Poisson mixture model (FZIP), and Poisson cluster weighted model (PCWM) on the data. Based on the confusion matrix, ZIPCWM achieves $97.4\%$ classification power, PCWM achieves $67.30\%$, while FZIP has the worst classification performance. In conclusion, ZIPCWM outperforms both PCWM and FZIP models.
	\bigskip
	\par \noindent \textbf{keywords:} Cluster-weighted model, Class Imbalance, Zero-inflated Poisson regression models, Poisson regression models, Fixed Zero-inflated Poisson model
\end{abstract}	

\section{Introduction}
Finite mixture of Poisson regression models with constant weights parameters have been developed by \cite{wedelatal1993}; \cite{branandrose1994}; \cite{wangetal1996}; and \cite{alfpandtrovato2004}. \cite{wangetal1998} incorporated covariates in the weight parameters of finite mixture Poisson regression models, treating the covariates as the concomitant variables. To account for different covariates with count response variables, \cite{Ingrassiaetal2015} proposed a generalized linear mixed cluster-weighted model where the response variable is allowed to follow the Poison distribution.   
As an alternative to Poisson regression model for handling overdispersion in data, a Negative binomial (NB) regression model can be used.
The count variable of interest may contain more zeros than expected under a Poisson model, which is observed in many applications. Inflated zeros can cause instability in a predictive model. Class imbalance also can be a consequence of zero inflation in a response variable we wish to predict therefore causing irregularities in the prediction accuracy and result in overfitting. Zero-Inflated Poisson (ZIP) regression model has been proposed with an application to defects in manufacturing (\cite{lambert1992}). The ZIP distribution is a mixture of a Poisson distribution and a degenerate at zero. This regression setting allows for the covariates in both Poisson mean and weight parameter. Furthermore, in a situation where overdispersion takes precedence, a zero-inflated negative binomial (ZINB) regression model can a better fit. However, if a population has excess zeros and several sub-populations in non-zero counts, a single component of the ZINB regression model may fail to capture the excess or sufficient to describe the non-zero counts. \par In this paper, we wish to address the problem of class imbalance by proposing a zero-inflated Poisson cluster weighted model that is capable of handling zero-inflation in data. We also wish to investigate further the effect of mixed covariates in the zero-inflated Poisson cluster weighted models. On this proposed model, we study the Expectation Maximization algorithm via iteratively re-weighted least squares (IRLS). At the E step, we follow the optimization conventionally. While at the M step, we used IRLS for the Poisson distribution. We compared the proposed model to the two existing models such as Poisson CWM (\cite{Ingrassiaetal2015}) and Zero-inflated Poisson regression model (\cite{hwa2014}). The paper is organized as follows. We describe briefly mixed continuous and categorical variables in Section \eqref{sec:2}. Section \eqref{sec:3} gives the details of the proposed model. We discuss the identifiability condition for the proposed model in Section \eqref{sec:4a}. In Section \eqref{sec:4}. We derived the EM algorithm via Iteratively rewieghted least squares for the proposed model. An extensive simulation study is carried out in section \eqref{sec:5}. Section \eqref{sec:6} demonstrates real data application of the model. Finally we conclude with a discussion in Section \eqref{sec:7}.
\section{Mixed Continuous and Categorical Variables}\label{sec:2}
We consider now the problem of fitting a family of cluster weighted model 
\begin{equation}
	p(\vec{x}, y) = \mathlarger \sum_{g=1}^{G} = \pi_g p(y|\vec{x},\mathcal{D}_g) p(\vec{x}|\mathcal{D}_g)
	\label{6.1}
\end{equation}
to some pair of data $(\mathbf{X}',Y)'$ where, $\mathbf{X}$ is a matrix of covariates, and $Y$ is a random variable defined on some space $\mathcal{D}$ where some of the feature variables are categorical. Basically, we can assume that the categorical variables are independent of each other and of the continuous variables which can be taken to have any continuous distribution such as the multivariate normal distribution. This idea of mixed covariates has come into prominence more recently due to its appearance in the graphical modeling of mixed variables known as the conditional Gaussian distribution model (\cite{whittaker1990}; \cite{coxandwermuth1992}). The location model has been used for fitting the mixture models to mixed categorical and continuous variables, (\cite{jorgensenandhunt1996}; \cite{lawrenceandkrzanowski1996}; and \cite{huntandjorgensen1999}). 
\section{The Zero-Inflated Poisson CWM}\label{sec:3}
Suppose that $\mathbf{X}$ can be decomposed as $(\mathbf{Q}', \mathbf{W})'$, where $\mathbf{Q}$ is a $q$-variate vector of continuous covariates and $\mathbf{W}$ is a $p$-variate vector of categorical variable respectively, being $d = q + p$. In this case, $\mathcal{X} = \mathbf{R}^{p\times \{1,...,r_1\},\times,...,\times \{1,...,r_p\}}$. With the location model, the $\mathbf{W}$ categorical variables are replaced by a single multinomial random variable $\mathbf{W}_i$ with $p$ cells. Let $Y$ be a count response variable, then the probability of observing zeros is $\pi_1$, and the probability of observing Poisson cluster weighted model is $1-\pi_1$.  The ZIP cluster weighted model is given as follows:
\begin{equation}
p(\mathbf{x},y;\mathbf{\Theta}) = 
\pi_1 I_{(y=0)} + \mathlarger \sum_{g = 2}^{G} p\big(y\big|\mathbf{x};\vec{\zeta}_g\big) p\big(\mathbf{x};\vec{\psi}_g\big)\pi_g, 
\label{6.2}
\end{equation}
where Equation \eqref{6.2} is decomposed into categorical and continuous covariates as follows;
\begin{equation}
= \pi_1 I_{(y=0)}+ \mathlarger \sum_{g = 2}^{G} p\big(y\big|\mathbf{x};\vec{\zeta}_g\big) p\big(\mathbf{q};\vec{\psi}^*_g\big)p\big(\mathbf{w};\vec{\psi}^{**}_g\big)\pi_g,
\label{6.3}
\end{equation}
since $y = 0$, and its mean $\mu_{i1}(\mathbf{x};\vec{\beta}_1) = 0$, then $p(y|\mathbf{x};\vec{\zeta}_1) = 1$. 
If we assume that the response variable contains many zeroes, then $p(y|\mathbf{x};\zeta_g)$ is modeled as a Zero-Inflated Poisson regression, $p(\mathbf{w}|\vec{\psi}^{**}_g)$ follows a multinomial distribution and $p(\mathbf{q}|\vec{\psi}^*_g)$ follows a Gaussian distribution, where $G$ is the number of mixing components,
$\pi_g$ is the mixing weight of component $g$ such that $0 < \pi_g < 1$, $g = 1, ..., G$ and $\sum_{g=1}^{G} \pi_g = 1$ by the decomposition $\pi_1 = 1-\sum_{g=2}^{G} \pi_g$. The weight $\pi_1$ determines the proportion of excess zeros compared with an ordinary Poisson mixture model determined by $1-\pi_1$. Assuming that $Y$ takes values in $\mathcal{Y}$ and that the conditional density $Y|\mathbf{x},\mathcal{D}_g$ is Poisson with parameter $\mu_{ig}(\mathbf{x};\vec{\beta}_g)$; that is, $Y|\mathbf{x},\mathcal{D}_g\sim P[\mu_{ig}(\mathbf{x};\vec{\beta}_g)]$. We allow the mean of the response variable to depend on the covariates $\mu_{ig}(\mathbf{x};\mathbf{\beta}_g)$ using the following regression models that
\begin{equation}
	\mu_{ig}(\mathbf{x};\vec{\beta}_g) = \exp(\mathbf{x}_i'\vec{\beta}_{g}), \hspace*{.1in} i = 1,...,N, g = 2,...,G
	\label{6.5}
\end{equation}
In this case,
\begin{equation}
	p(y|\mathbf{x};\vec{\beta}_g) = \exp[-\mu_{ig}(\mathbf{x};\vec{\beta}_g)]\frac{[-\mu_{ig}(\mathbf{x};\vec{\beta}_g)]^y}{y!}
	\label{6.6}
\end{equation}
The mixing weights $\lbrace\pi_g\rbrace_{g=1}^G$ can be treated as the multinomial logit of $\pi_g$ to be linear function of covariates which is commonly known in the literature as a concomitant variable.
\begin{equation}
	\pi_{ig}(\mathbf{v}_i,\gamma) = \frac{\exp(\mathbf{v}_i\gamma_g)}{\mathlarger\sum_{g=1}^{G}\exp(\mathbf{v}_i\gamma_g)}
	\label{6.7} 
\end{equation}
In this case, $\mathbf{x}_i = (x_{i1},..., x_{ip})$ and $\mathbf{v}_i = (v_{i1},..., v_{ir})$ are $1\times p$ and $1\times q$ rows vectors of covariates (including  an intercept), respectively. They can be the same or have nothing in common. The regression coefficient for the $g$th component are $\beta_g$ and $\gamma_g$ which are vectors of $p\times 1$ and $q\times 1$ respectively. We note that the mixing probability of the first component $\pi_{i1}(\mathbf{v}_i, \gamma)$ is taken to be the baseline for the multinomial logistic model.
To handle response variable with excess zeros, we assume that the density of $p(y|\mathbf{x};\vec{\eta}_g)$ belong to exponential family. It is a common practice that the exponential family is strictly related to the generalized linear models with a monotone and differential link function $f(.)$ that makes the expected value $\mu_g$, of $Y|\mathcal{D}_g$ depend on the covariate $\mathbf{X}$ through the linear combination $f(\mu_g) = \mathbf{x}'\vec{\beta}_g$. Note here that we have not used $\vec{\beta}_g$ explicitly, so it contains the coefficient and $\mathbf{X}$ will have extra column of ones. The distribution of $Y|\mathcal{D}_g$ is denoted by $p(y|\mathbf{x};\vec{\beta}_g)$.
The term $p(\mathbf{q};\vec{\psi}^*_g)$ in Equation \eqref{6.3} is modeled as a $q$-variate Gaussian density with mean $\vec{\mu}_g$ and covariance matrix $\vec{\Sigma}_g$, i.e., $p(\mathbf{q}|\vec{\psi}^*_g) = \phi(\mathbf{q};\vec{\mu}_g,\vec{\Sigma}_g)$. 
With respect to the term $p(\mathbf{w};\vec{\psi}^{**}_g)$ in Equation \eqref{6.3}, we assume that each categorical covariate can be taken to be a binary vector $\mathbf{w}^k = (w^{k1},...,w^{kr_k})'$, where $w^{ks} = 1$ if $r_k$ is equal to the value $s$, with $s \in \{1,...,r_k\}$, and $w^{ks} = 0$ otherwise. Furthermore, assume that $q$ categorical covariates are independent of each other. Then, 
\begin{equation}
p(w;\vec{\alpha}_g) = \mathlarger \prod_{k=1}^{p}\mathlarger \prod_{s=1}^{r_k} (\alpha_{gks})^{w^{ks}}, 
\label{6.8}
\end{equation}
where $g = 1,..., G$, $\vec{\alpha}_g = (\vec{\alpha}_{g1}',...,\vec{\alpha}'_{gp})'$ and $\vec{\alpha}_{gk} = (\alpha_{gk1},...,\alpha_{gkr_p})'$. We take the density of $p(\mathbf{w};\vec{\alpha}_g)$ in Equation \eqref{6.8} to be a multinomial distribution of parameters $\vec{\alpha}_{gk}$, where $k = 1,...,p$ and $\sum_{s=1}^{r_k} \alpha_{gks} = 1$ with constraint $\alpha_{gks} > 0$.
The ZIPCWM over all observation can be formulated as follows:
\begin{equation}
p(\vec{x},y;\mathbf{\Theta}) = 
\pi_1 I_{(y=0)} + \mathlarger \sum_{g = 2}^{G} Pois\big(y\big|\vec{x};\vec{\beta}_g\big) \phi\big(\vec{q};\vec{\mu}_g,\vec{\Sigma}_g\big)p\big(\vec{w};\vec{\alpha}_g\big)\pi_g
\label{6.9}
\end{equation}
where $I_{(.)}$ is an indicator function that outputs $1$ when the specified condition is satisfied and $0$ otherwise, and Pois(.) denotes the probability mass function of $y_i$ and $x_i$ with a mean of $\mu_{ig}(\mathbf{x}_i;\vec{\beta}_g)$.
\subsection{Related Mixture Model}
This section highlights the special cases of the proposed model. We show a list of special cases of ZIPCWM below;
\subsubsection{Poisson CWM \cite{Ingrassiaetal2015}}
Let $p(\mathbf{x},y;\mathbf{\Theta})$ in Equation \eqref{6.9} be a ZIPCWM. Assuming that $Y$ takes values $\mathrm{y} = \mathcal{N}$, $Y|\mathbf{x};\mathcal{D}_g$ is Poisson regression model with parameter $\mu_g(\mathbf{x};\vec{\beta}_g)$ and the $\mathbf{X}|\mathcal{D}_g$ with parameter $(\vec{\mu}_g, \vec{\Sigma}_g)$. If there is no account of excess zeros, then $Y|\mathbf{x},\mathcal{D}_g$ is modeled with Poisson regression and the $X|\mathcal{D}_g$ is taken to be the Gaussian distribution, then ZIPCWM reduces to Poisson CWM as follows;
\begin{equation}
	p(\mathbf{x},y;\mathbf{\Theta}) = 
	\mathlarger \sum_{g = 1}^{G} Pois\big(y\big|\mathbf{x};\vec{\beta}_g\big) \phi\big(\mathbf{q};\vec{\mu}_g,\vec{\Sigma}_g\big)p\big(\mathbf{w};\vec{\alpha}_g\big)\pi_g
	\label{6.10}
\end{equation}
\subsubsection{ZIP Regression mixture model \cite{hwa2014}}
If $p(\mathbf{x}) = 1$ in Equation \eqref{6.2}, then ZIPCWM reduces to GZIP mixture distribution if the mixing weight depends on a concomitant variables and FZIP if the mixing weight is fixed.
Generalized ZIP model:
\begin{equation}
	p(y;\mathbf{\Theta}) = 
	\pi_{i1}(\mathbf{v}_i,\gamma) I_{(y=0)} + \mathlarger \sum_{g = 2}^{G} Pois\big(y\big|\mathbf{x};\vec{\beta}_g\big)\pi_{ig}(\mathbf{v}_i,\gamma)
	\label{6.11}
\end{equation}
and the fixed ZIP model is as follows:
\begin{equation}
	p(y;\mathbf{\Theta}) = 
	\pi_1 I_{(y=0)} + \mathlarger \sum_{g = 2}^{G} Pois\big(y\big|\mathbf{x};\vec{\beta}_g\big)\pi_g
	\label{6.12}
\end{equation}
\subsubsection{ZIP distribution \cite{lambert1992}:}
If $p(\mathbf{x}) = 1$ and $G = 2$ in Equation \eqref{6.2}, then ZIPCWM reduces to ZIP distribution.
\subsubsection{Standard Poisson mixture model:} 
If both $\pi_g$ and $\mu_g(\mathbf{x}_i,\beta_g)$ are constant functions and $p(\mathbf{x}) = 1$, ZIPCWM reduces to the standard Poisson mixture model, denoted by 
\begin{equation}
	p(Y=y_i) = \sum_{g=1}^{G} \pi_g Pois(y_i|\lambda_g)
	\label{6.13}
\end{equation}
where the constancy of $\mu_{ig}(\mathbf{x}_i,\beta_g)$ is taken to be $\lambda_g$. It should be noted that we have used $\lambda_g$ to preserve the original parameter notation of the standard Poisson mixture model.
In the following, we present the derivative of an estimation method based on the EM algorithm for the ZIPCWM regression model described in Equation \eqref{6.3}.
\section{Identifiability}\label{sec:4a}
To reliably estimate the parameters of Equation \eqref{6.9}, we require that the ZIPCWM be identifiable, that is, two sets of parameters which do not agree after permutation cannot produce the same mixture distribution. \cite{teicher1961} proves that the class of finite mixtures of Poisson distribution is identifiable without covariates. Similarly, \cite{follmannandlambert1991} give sufficient conditions for identifiability, and \cite{wangetal1996} extend the definition of identifiability of finite Poisson mixtures with covariates. We provide the condition for identifiability of ZIPCWM as follows;
Consider the collection of probability models $\{p(\mathbf{x}_1,y_1,\vec{\Theta}),...,p(\mathbf{x}_N,y_N;\vec{\Theta})\}$, with a restriction that $\pi_1 <...< \pi_g$, sample space $\mathcal{Y}_1,...,\mathcal{Y}_N$, parameter space $\vec{\Theta}$, and fixed covariates vectors $x_1,...,x_N$ that is decomposed to categorical $w$ and continuous variables $q$, where $w_i \in \mathcal{R}^p$ and $q_i \in \mathcal{R}^d$ for $i = 1,...,N$. The collection of probability model is \textit{identifiable} if for $(\vec{\Omega}, \vec{\pi}, \mathbf{k}, \mathbf{h})$, $(\vec{\Omega}^*, \vec{\pi}^*, \mathbf{k}^*, \mathbf{h}^*)$ $\in \vec{\Theta}$,
\begin{equation}
	p(\mathbf{x},y,\vec{\Omega}, \vec{\pi}, \mathbf{k}, \mathbf{h}) = p(\mathbf{x},y,\vec{\Omega}^*, \vec{\pi}^*, \mathbf{k}^*, \mathbf{h}^*)
	\label{equ:5.100}
\end{equation}
for all $y_i \in \mathcal{Y}_i, i=1,...,N$, implies that $(\vec{\Omega}, \vec{\pi}, \mathbf{k}, \mathbf{h}) = (\vec{\Omega}^*, \vec{\pi}^*, \mathbf{k}^*, \mathbf{h}^*)$. \par \noindent We note here that the restriction on the mixing probability is a sufficient condition for label switching problems and it means that two models are equivalent if they agree up to permutation of parameters. \par \noindent We now provide a sufficient condition for identifiability. Suppose that $(\vec{\Omega}, \vec{\pi}, \mathbf{k}, \mathbf{h})$, $(\vec{\Omega}^*, \vec{\pi}^*, \mathbf{k}^*, \vec{h}^*)$ satisfy Equation \ref{equ:5.100}. 
It then implies that
\begin{equation}
	\pi_1 I_{(y=0)} + \mathlarger \sum_{g = 2}^{G} \text{Pois}\big(y\big|\mathbf{x};\vec{\beta}_g\big) \phi\big(\mathbf{q};\vec{\mu}_g,\vec{\Sigma}_g\big)p\big(\mathbf{w};\vec{\alpha}_g\big)\pi_g = 
	\notag
\end{equation} 
\begin{equation}
	\pi_1^* I_{(y=0)} + \mathlarger \sum_{c = 2}^{C} \text{Pois}\big(y\big|\mathbf{x};\vec{\beta}_c^*\big) \phi\big(\mathbf{q};\vec{\mu}_c^*,\vec{\Sigma}_c^*\big)p\big(\mathbf{w};\vec{\alpha}_c^*\big)\pi_c^*
	\label{equ:5.110}
\end{equation}
for each $i$ and $y_i \in \mathcal{Y}_i, i=1,...,N$. Teicher's and Hennig's results imply that 
\begin{equation}
	G=C, \hspace{0.05in}\pi_g=\pi^*_c, \hspace{0.05in} \vec{\beta}_g=\vec{\beta}_c^*, \hspace*{0.05in} \vec{\mu}_g^*=\vec{\mu}^*_c,
	\notag
\end{equation} 
\begin{equation}
	\vec{\Sigma}_g=\vec{\Sigma}^*_c, \hspace{0.05in}\text{and}\hspace{0.05in} \vec{\alpha}_g=\vec{\alpha}_c^*
	\notag
\end{equation} 
for $i=1,...,N$ $g=1,...,G$, and $c=1,...,C$. By definition of ZIPCWM, we obtain 
\begin{equation}
	\exp(\mathbf{x}_i'\vec{\beta}_{g}) = \exp(\mathbf{x}_i'\vec{\beta}^*_{c}),
	\label{equ:5.120}
\end{equation}
Equation \ref{equ:5.120} means 
\begin{equation}
	(\vec{\beta}_g - \vec{\beta}^*_c)'\mathbf{x}_i=0, \hspace*{0.05in} \text{for} \hspace*{0.05in} i = 1,...,N.
\end{equation}
Then, we say 
\begin{equation}
	\text{Pois}\big(y\big|\mathbf{x};\vec{\beta}_g\big) = \text{Pois}\big(y\big|\mathbf{x};\vec{\beta}_c^*\big), \phi\big(\mathbf{q};\vec{\mu}_g,\vec{\Sigma}_g\big) = \phi\big(\mathbf{q};\vec{\mu}_c^*,\vec{\Sigma}_c^*\big),
	\notag
\end{equation}
\begin{equation}
	p\big(\mathbf{w};\vec{\alpha}_g\big) = p\big(\mathbf{w};\vec{\alpha}_c^*\big),
	\pi_g = \pi_c^*.
	\label{equ:5.150}	
\end{equation}
Hence a sufficient condition for identifiability is that $\mathbf{X}$ is a full rank matrix, where $\mathbf{X}={\mathbf{x}_1,...,\mathbf{x}_N}$.
\section{Model Estimation by EM Algorithm}\label{sec:4}
Let $(\mathbf{x}'_i,y_i),...,(\mathbf{x}'_n,y_n)'$ be a sample of $n$ independent observation pairs drawn from model Equation \eqref{6.9}. The corresponding likelihood, for a fixed number of component $G$ is given by 
\begin{equation}
	L(\mathbf{\Theta}) = \mathlarger \prod_{i=1}^{n} p(\mathbf{x}_i,y_i;\mathbf{\Theta}) = \mathlarger \prod_{i=1}^{n} \bigg[\pi_1 I_{(y=0)} + \mathlarger \sum_{g = 2}^{G} Pois\big(y\big|\mathbf{x};\vec{\beta}_g\big) \phi\big(\mathbf{q};\vec{\mu}_g,\vec{\Sigma}_g\big)p\big(\mathbf{w};\vec{\alpha}_g\big)\pi_g\bigg]
	\label{6.14}
\end{equation}
Here, we assume the number of $G$ is fixed and known a priori and $\mathbf{z}_i = (z_{i1},...,z_{iG})'$ be the latent vector of component indicator variables, where $z_{ig} = 1$ if ith subject $(\mathbf{x}'_i,y_i)'$ comes from $\mathcal{D}_g$ belongs to the $gth$ latent group and $z_{ig} = 0$ otherwise. By assumption, each belongs to one of the $G$ unobservable components therefore, considered missing or incomplete. Using a multinomial distribution for the unobserved vector $\mathbf{z}_i$,where $\{(\mathbf{x}'_i,y_i,\mathbf{z}'_i)';i = 1,...,n\}$ is th complete data. The the complete-data log-likelihood can be written as 
\begin{equation}
L_{\mathbf{c}}(\mathbf{\Theta}) = \mathlarger \prod_{i=1}^{n}\bigg[\pi_1 I_{(y=0)} + \mathlarger \prod_{g = 2}^{G} Pois\big(y\big|\mathbf{x};\vec{\beta}_g\big) \phi\big(\mathbf{q};\vec{\mu}_g,\vec{\Sigma}_g\big)p\big(\mathbf{w};\vec{\alpha}_g\big)\pi_g\bigg]^{z_{ig}}
\label{6.15}
\end{equation}
The corresponding complete-data log-likelihood by taking the logarithm of Equation \eqref{6.14} can be written as
\begin{equation}
l_{\mathbf{c}}(\mathbf{\Theta}|y, \vec{\omega}, \mathbf{x}) = \mathlarger\sum_{i=1}^{N} z_{i1}\ln\pi_1I_{y_i=0} + \mathlarger\sum_{i=1}^{N}\mathlarger \sum_{g=2}^{G}z_{ig}\bigg[\ln\pi_g + \ln Pois\big(y_i|\mu_{ig}(\mathbf{x}_i,\vec{\beta}_g)\big)
\notag
\end{equation}
\begin{equation}
+ \ln \phi\big(\mathbf{q}_i|\vec{\mu}_g,\vec{\Sigma}_g\big) + \ln p\big(\mathbf{w}_i;\vec{\alpha}_g\big)\bigg]
\label{6.16}
\end{equation}
where $\pi_1 = 1 - \sum_{g=2}^{G}\pi_g$ and $\mathbf{\Theta} = (\vec{\mu}_g, \vec{\Sigma_g}, \vec{\beta}_g)$ for $g = 2,...G$ is the set of model parameters to be estimated.
\subsection{E-step}
Using the current estimates $\mathbf{\Theta}^{(t)}$, we compute the probability $z^{(t)}_{ig}$ that the subject $i$ comes from $gth$ component of the mixture:
\begin{equation}
E(z_{ig}|y_i,\vec{\omega}_i, \mathbf{x}_i, \mathbf{\Theta}^{(t)}) = {z}^{(t)}_{ig} = \frac{ \pi_g Pois\bigg(y_i|\mu_{ig}(\mathbf{x}_i,\vec{\beta}_g^{(t)})\bigg)\phi\bigg(\mathbf{x}_i|\vec{\mu}_g^{(t)},\vec{\Sigma}_g^{(t)}\bigg)}{p\bigg(\mathbf{x}_i,y_i;\vec{\Theta}^{(t)}\bigg)}
\label{6.17}
\end{equation}
which correspond to the posterior probability that the unlabeled observation $(\mathbf{x}'_i,y_i)'$ belong to the $g$th component of the mixture, using the current fit $\mathbf{\Theta}^{(t)}$ for $\mathbf{\Theta}$ and ${z}^{(t)}_{i1} = 1 - \sum_{g=2}^{G}{z}^{(t)}_{ig}.$
\subsection{M-step} 
At this step, we obtain the $Q(.)$ with respect to $\mathbf{\Theta}^{(t+1)}$ where $t =0, 1, \ldots$, The conditional expectation of $l_c(\mathbf{\Theta})$ given the observed data , say $Q(\mathbf{\Theta};\mathbf{\Theta}^{(t)})$ is maximized with respect to $\mathbf{\Theta}$. The $z_{ig}$ are simply replaced by the current expectations $z_{ig}^{(t)}$ yielding,
\begin{equation}
Q\bigg(\mathbf{\Theta};\mathbf{\Theta}^{(t)}\bigg) = \mathlarger\sum_{i=1}^{N} z^{(t)}_{i1}\ln\pi_1 I_{y_i=0} + \mathlarger\sum_{i=1}^{N}\mathlarger \sum_{g=2}^{G}z^{(t)}_{ig}\ln\pi_g + \mathlarger \sum_{i=1}^{N} \mathlarger \sum_{g=2}^{G}z^{(t)}_{ig}\ln p\big(\mathbf{w}_i;\vec{\alpha}_g\big)
\notag
\end{equation}
\begin{equation}
	+ \mathlarger \sum_{i=1}^{N} \mathlarger \sum_{g=2}^{G}z^{(t)}_{ig}\ln Pois\big(y_i|\mu_{ig}(\mathbf{x}_i,\vec{\beta}_g)\big) + \mathlarger \sum_{i=1}^{N} \mathlarger \sum_{g=2}^{G}z^{(t)}_{ig}\ln\bigg(\phi\big(\vec{q}_i|\vec{\mu}_g,\vec{\Sigma}_g\big)\bigg) 
\label{6.18}
\end{equation}
\subsubsection{Parameter Estimation of $Y$}
We maximize Equation \eqref{6.18} independently of the $G$ expression as the four terms on the right of equation \ref{6.18} have zero cross-derivatives, i.e.
\begin{equation}
 \frac{\partial}{\partial \vec{\beta}_g} \mathlarger \sum_{i=1}^{N} z^{(t)}_{ig}\ln \bigg(Pois\big(y_i|\mu_{ig}(\mathbf{x}_i,\vec{\beta}_g)\big)\bigg)
\label{6.19}
\end{equation}
\begin{equation}
   \frac{\partial}{\partial \vec{\beta}_g}\mathlarger \sum_{i=1}^{n} z^{(t)}_{ig} \log \frac{\bigg(\mu_{ig}(\mathbf{x}_i;\vec{\beta}_g)\bigg)^{y_i}\exp \bigg(-\mu_{ig}(\mathbf{x}_i;\vec{\beta}_g)\bigg)}{y_i!}
	\label{6.20}
\end{equation}
\begin{equation}
	\frac{\partial}{\partial \vec{\beta}_g} \mathlarger \sum_{i=1}^{n} z_{ig}^{(t)} \bigg[y_i\log \mu_{ig}(\mathbf{x}_i;\vec{\beta}_g) - \mu_{ig}(\mathbf{x}_i;\vec{\beta}_g) - \log y_i!\bigg]
	\label{6.21}
\end{equation}
\begin{equation}
	\frac{\partial}{\partial \vec{\beta}_g} \mu_{ig}(\mathbf{x}_i;\vec{\beta}_g) = \mathbf{x}_i\exp(\mathbf{x}_i'\vec{\beta}_g) = \mathbf{x}_i \mu_{ig}(\mathbf{x}_i;\vec{\beta}_g)
	\label{6.22}
\end{equation}
\begin{equation}
\frac{\partial}{\partial \vec{\beta}_g} \mathlarger \sum_{i=1}^{n} z_{ig}^{(t)} \bigg[y_i\log \mu_{ig}(\mathbf{x}_i;\vec{\beta}_g) - \mu_{ig}(\mathbf{x}_i;\vec{\beta}_g) - \log y_i!\bigg]
\label{6.23}
\end{equation}
From Equation \eqref{6.21}, we have the following 
\begin{equation}
 \mathlarger \sum_{i=1}^{n} z_{ig}^{(t)}\mathbf{x}_i\mu_{ig}\big(\mathbf{x}_i;\vec{\beta}_g\big) \bigg(\frac{y_i}{\mu_{ig}(\mathbf{x}_i;\vec{\beta}_g)} - \frac{\mu_{ig}(\mathbf{x}_i;\vec{\beta}_g)}{\mu_{ig}(\mathbf{x}_i;\vec{\beta}_g)}\bigg) = 0
\label{6.24}
\end{equation}
\begin{equation}
S(\vec{\beta}^{(t)}_g) = \mathlarger \sum_{i=1}^{n} z_{ig}^{(t)}\mathbf{x}_i\bigg(y_i - \mu_{ig}(\mathbf{x}_i;\vec{\beta}_g)\bigg)
\label{6.25}
\end{equation}
Using the expression $\vec{\beta}_g^{(t+1)} = \vec{\beta}_g^{(t)} + \big[I(\vec{\beta}_g^{(t)})\big]^{-1} S(\vec{\beta}_g^{(t)})$, where $I(\vec{\beta}_g^{(t)})$ is the Fisher Information matrix and $S(\vec{\beta}_g^{(t)})$ is the score function obtained in Equation \eqref{6.25}. So,
\begin{equation}
	I(\vec{\beta}_g^{(t)}) = -\frac{\partial}{\partial} S(\vec{\beta}_g^{(t)}) = \mathlarger\sum_{i=1}^{n} z^{(t)}_{ig} \mu_{ig}(\mathbf{x}_i;\vec{\beta}_g)\mathbf{x}'_i\mathbf{x}_i 
	\label{6.26}
\end{equation}
\begin{equation}
	\vec{\beta}_g^{(t+1)} = \vec{\beta}_g^{(t)} + \bigg(\mathlarger\sum_{i=1}^{n} z^{(t)}_{ig} \mu_{ig}\big(\mathbf{x}_i;\vec{\beta}_g\big)\mathbf{x}'_i\mathbf{x}_i\bigg)^{-1}\bigg(\mathlarger \sum_{i=1}^{n} z_{ig}^{(t)}\mathbf{x}_i\big(y_i - \mu_{ig}(\mathbf{x}_i;\vec{\beta}_g)\big)\bigg)
	\label{6.27}
\end{equation}
finally Equation \eqref{6.27} becomes
\begin{equation}
	\vec{\beta}_g^{(t+1)} = \bigg(\mathlarger\sum_{i=1}^{n} \mathbf{x}'_is_{ig}\mathbf{x}_i\bigg)^{-1}\bigg(\mathlarger \sum_{i=1}^{n} \mathbf{x}_i's_{ig}\delta_{ig}^{(t)}\bigg)
	\label{6.28}
\end{equation}
where $\delta^{(t)} = \mathbf{x}'_i\vec{\beta}_g + \delta_{ig}^*$ with $\delta_{ig}^* = \big(y_i - \mu_{ig}(\mathbf{x}_i;\vec{\beta}_g)\big)\big/ \mu_{ig}(\mathbf{x}_i;\vec{\beta}_g)$ and $s_{ig} = z^{(t)}_{ig}\mu^{(t)}_{ig}(\mathbf{x}_i;\vec{\beta}_g)$
\subsubsection{Parameter Estimation for Mixing Weights}
The maximum of $Q(\mathbf{\Theta})$ with respect to $\vec{\pi}$, subject to the constraints on these parameters. is achieved by maximizing the augmented function
\begin{equation}
	= \mathlarger\sum_{i=1}^{N} z^{(t)}_{i1}\ln \pi_1 - \gamma_1 \biggl(\pi_1 - 1 \biggr)I_{y_i=0} + \mathlarger\sum_{i=1}^{N} \mathlarger \sum_{g=2}^{G}z^{(t)}_{ig}\ln\pi_g + \gamma \biggl(\mathlarger \sum_{g=2}^{G} \pi_g - 1 \biggr)
	\label{6.29}
\end{equation}
where $\gamma_1$ is a Lagrangian multiplier. Setting the derivative in Equation \eqref{6.29} with respect to $\pi_1$ for a degenerate distribution and $\pi_g$ for a Poisson cluster weighted models equal to zero and solving yields
\begin{equation}
\pi_1^{(t+1)} = \mathlarger \sum_{i=1}^{n} z^{(t+1)}_{i1} \bigg/ n
\notag
\end{equation}
\begin{equation}
\pi_g^{(t+1)} = \mathlarger \sum_{i=1}^{n} z^{(t+1)}_{ig} \bigg/ n
\label{6.30}
\end{equation}
where $g = 2,\ldots, G$.
\subsubsection{Parameter Related to q}
Maximizing Equation \eqref{6.18} with respect to $\vec{\mu}_g$ and $\vec{\Sigma}_g$, $g = 1,...,G$, is equivalent to maximizing independently each of the $G$ expression
\begin{equation}
	\mathlarger \sum_{i=1}^{n}z^{(t)}_{ig}\ln\bigg(\phi\big(\mathbf{q}_i|\vec{\mu}_g,\vec{\Sigma}_g\big)\bigg)
	\label{6.50}
\end{equation}
we obtain
\begin{equation}
	\vec{\mu}_g^{(t+1)} = \mathlarger \sum_{i=1}^{n}z^{(t)}_{(ig)}\mathbf{q}_i\bigg/\mathlarger \sum_{i=1}^{n}z^{(t)}_{(ig)}
	\label{6.51}
\end{equation}
and
\begin{equation}
\vec{\Sigma}_g^{(t+1)} = \mathlarger \sum_{i=1}^{n} z_{ig}^{(t)}\big(\mathbf{q}_i - \vec{\mu}_g^{(t+1)}\big)\big(\mathbf{q}_i - \vec{\mu}_g^{(t+1)}\big)'\Bigg /\mathlarger \sum_{i=1}^{n}z_{ig}^{(t)}
\label{6.52}
\end{equation}
\subsubsection{Parameters Related to W}
Maximizing $Q(\mathbf{\Theta};\mathbf{\Theta}^{(t)})$ with respect to $\vec{\alpha}_g$, $g = 1,\ldots, G$ is equivalent to maximizing each of the $G$ expression considering the constraints
\begin{equation}
	\mathlarger \sum_{i=1}^{n} z_{ig}^{(t)} \ln p(\mathbf{w}_i;\vec{\alpha}_g) = \mathlarger \sum_{k=1}^{p} \mathlarger \sum_{i=1}^{n} z_{ig}^{(t)} \mathlarger \sum_{s=1}^{r_k} w^{ks}\ln \alpha_{gks}.
	\label{6.34}
\end{equation}
Using the Lagrangian multiplier, Equation \eqref{6.34} can be expanded as follows;
\begin{equation}
	\mathlarger \sum_{i=1}^{n} z_{ig}^{(t)} \mathlarger \sum_{s=1}^{r_k} w^{ks}\ln \alpha_{gks} - \gamma_2\bigg(\mathlarger \sum_{s=1}^{r_k}\alpha_{gks} - 1 \bigg),
	\label{6.35}
\end{equation} 
the $\gamma_2$ is used here as the Lagrangian multiplier. Setting the derivative of Equation \eqref{6.35} to zero
\begin{equation}
	\mathlarger \sum_{i=1}^{n} z_{ig}^{(t)} \frac{w^{ks}}{\alpha_{gks}} - \gamma_2 = 0, \implies
	\mathlarger \sum_{i=1}^{n} z_{ig}^{(t)} w^{ks}\bigg/\gamma_2 = \alpha_{gks},	
\label{6.36}
\end{equation}
we find $\gamma_2 = \sum_{i=1}^{n} z_{ig}^{(t)}$, then finally we have 
\begin{equation}
	\alpha_{gks}^{(t+1)} = \mathlarger \sum_{i=1}^{n} z_{ig}^{(t)} w^{ks} \bigg/ \sum_{i=1}^{n} z_{ig}^{(t)},	
	\label{6.37}
\end{equation}
\clearpage
\begin{figure}[H]
	\centering
	\begin{minipage}[b]{0.45\textwidth}
		\includegraphics[width = 3in, height=2in]{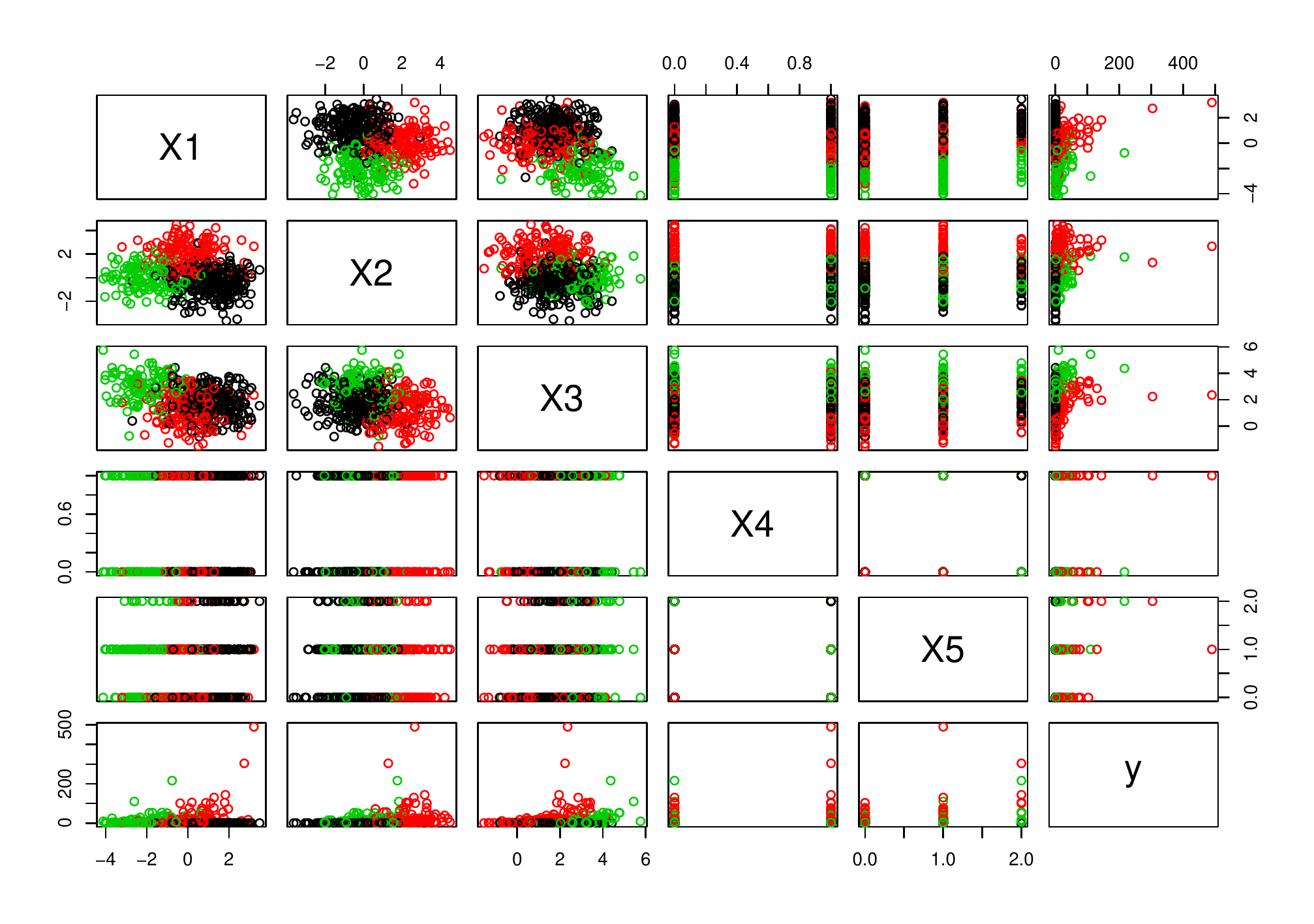}
		\centering\caption{\textit{The Visualization of the simulated data with sample size $500$}} 
		\label{fig:a}
	\end{minipage}
	\hfil
	\begin{minipage}[b]{0.45\textwidth}
		\includegraphics[width = 3in, height=2in]{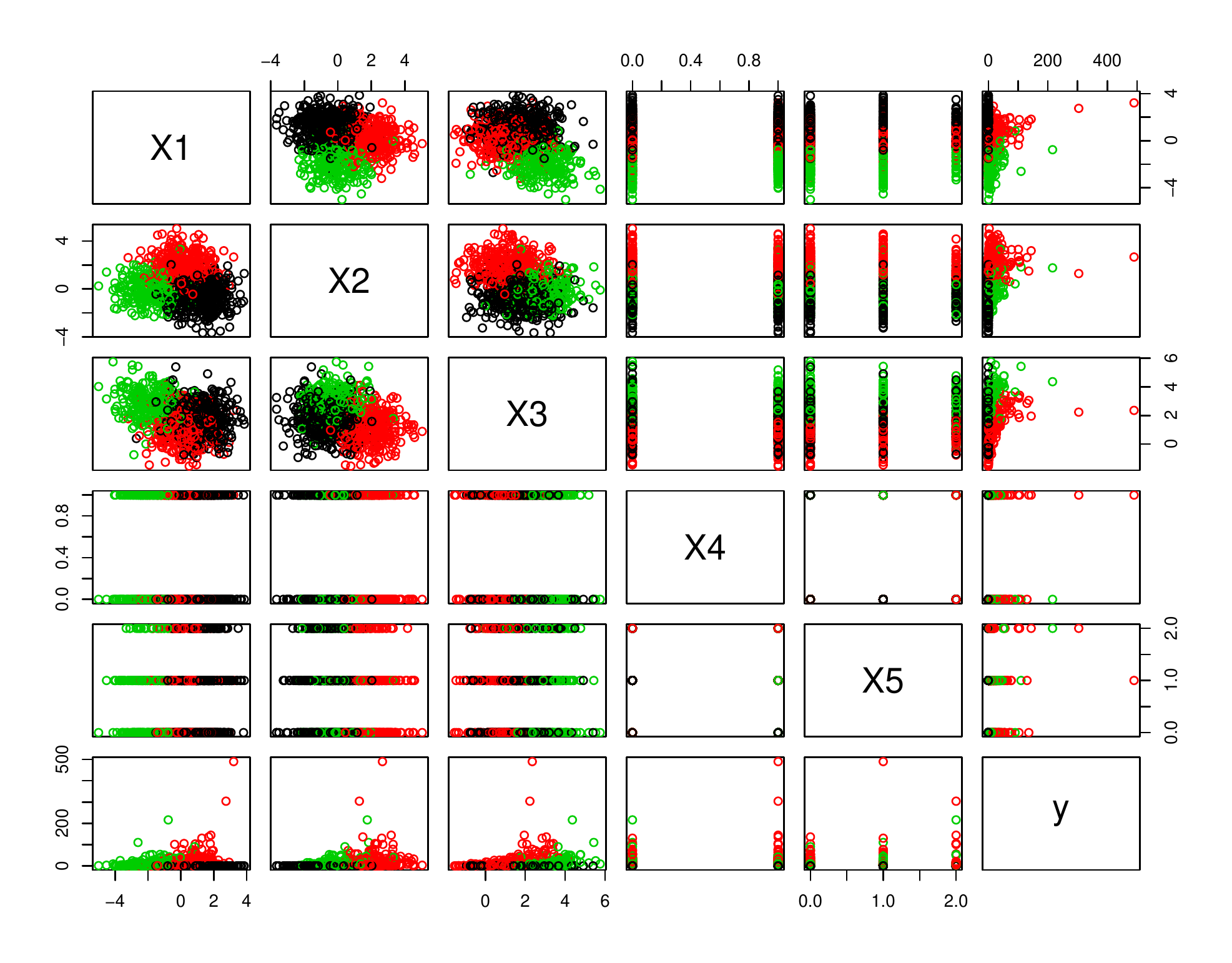}
		\caption{\textit{The Visualization of the simulated data with sample size $1000$}} 
		\label{fig:b}
	\end{minipage}
\end{figure}
\section{Simulation Study}\label{sec:5}
A simulation study was performed to evaluate the performance of the maximum likelihood estimates of the model obtained via EM algorithm. We generated samples of size $n$ from the following ZIPCWM with three components.
\begin{equation}
p(\mathbf{x},y;\mathbf{\Theta}) = 
\pi_1I_{(y=0)} + \mathlarger \sum_{g = 2}^{3} Pois\big(y\big|\mathbf{x};\vec{\beta}_g\big) \phi\big(\mathbf{q};\vec{\mu}_g,\vec{\Sigma}_g\big)p\big(\mathbf{w};\vec{\alpha}_g\big)\pi_g
\end{equation}
\par \noindent The log-link for the Poisson mean $\mu_{ik}$ is as follows:
\begin{equation}
	\log(\mu_{i2}(\mathbf{x}_i, \beta_2)) = \beta_{20} + \beta_{21}\mathbf{x}_i
	\notag
\end{equation}
\begin{equation}
	\log(\mu_{i3}(\mathbf{x}_i, \beta_3)) = \beta_{30} + \beta_{31}\mathbf{x}_i
	\label{6.38}
\end{equation}
where $\mathbf{x}_i$ in this case is the combination of $\mathbf{q}_i, \text{and} \hspace{0.06in}\mathbf{w}_i$ random variables. Figure \eqref{fig:a} and Figure \eqref{fig:b} present the visualization of the simulated data with sample sizes $500$ and $1000$ but we intentionally not include visualization for $n = 200$. We considered the distribution $\mathbf{q}|\mathcal{D}_g \in \mathcal{R}^d$ to follow a Gaussian distribution. Table \eqref{tab:1} shows the true values of the mean vectors, and the covariance matrices are the spherical covariance matrices.
For the finite discrete variables $w_1$ and $w_2$, we generated two different variables from binomial distribution where $p = 2$ discrete covariates and $r_1 = 2$ and $r_2 = 3$ levels with the probability of success $0.5$ and $1/3$ respectively.
The true values of the ZIPCWM regression coefficients are presented in Table \eqref{tab:2}. We selected the $\pi_{1} \approx 0.5$, $\pi_{2} \approx 0.3$, and $\pi_{3} \approx 0.2$ according to \cite{hwa2014}. We generated a random number $U$ from Uniform distribution. To generate samples from the above model for each subject $(i = 1,...,n)$, we adopted the following conditions; if $U$ is less than $\pi_{1}$, $Y_i$ takes the value $0$, where $\mu_{i1} = 0$. If $U$ is between $\pi_1$ and $\pi_1 + \pi_2$, then $Y_i$ is a draw from Pois($\mu_{i2}$). Otherwise, $Y_i$ is generated from Pois$(\mu_{i3})$.
\subsection{Result for Parameters Estimated}
Here, the ZIPCWM with three component is fitted to the simulated data of sizes $n = 200, 500$, and $1000$. We compute the misclassification based on the simulated data. We ran each simulation $10$ times as suggested by\cite{McandP2000} to avoid local maxima. However, we ran the simulated $10$ times and recorded the average values to avoid the biased choice of selecting the best result. Table \eqref{tab:1} shows the recovered estimates of the $\vec{\mu}_g$, the diagonal values for $\vec{\Sigma}_g$, and the mixing proportions $\vec{\pi_g}$. For estimated parameters of $\mu_1$ in component $1$ for $n = 200, 500$, and $n = 1000$, the values are omitted. Generally, in Table \eqref{tab:1}, we observe that the closeness of the parameter values is not independent of the sample size $n$. Also, in Table \eqref{tab:2}, considering the coefficients $\vec{\beta}_g$ where $g = 2,\ldots,G$. The parameter estimates for $n = 1000$ are closer to the true parameter values compared to the parameter estimates for $n = 200$ and $n = 500$. We investigate the ability of the proposed model to identify the number of 
\begin{table}[H]
	\centering\caption{True and Recovered values of the mean, sigma, and mixing weight\\ \centering for $n = 200, 500$ and $1000$}\vspace{0.4cm}
	\label{tab:1}
	\begin{tabular}{r@{\hspace{.2in}}r@{\hspace{.2in}}r@{\hspace{.1in}}r@{\hspace{.1in}}r@{\hspace{.1in}}r@{\hspace{.1in}}r@{\hspace{.1in}}r@{\hspace{.1in}}r@{\hspace{.1in}}rr}
		\hline
		&$n$&$g$ &$\mu_1$& $\mu_2$& $\mu_3$&$\sigma_{11}$&$\sigma_{22}$&$\sigma_{33}$&$\vec{\pi}$&\\ 
		\hline
		True&&$1$&-&$-$&$-$&$-$&-&-&$0.50$&\\
		&&$2$&$0.10$&$2.00$&$1.00$&$1.00$&$1.00$&$1.00$&$0.30$&\\ 
		&&$3$&$-2.00$&$0.00$&$3.00$&$1.00$&$1.00$&$1.00$&$0.20$&\\ 
		\hline
		Recovered&$200$&$1$&$-$&$-$&$-$&$-$&$-$&$-$&$0.54$&\\
		&&$2$&$-0.77$&$1.50$&$2.20$&$1.99$&$1.62$&$1.48$&$0.25$&\\
		&&$3$&$-1.07$&$0.92$&$2.40$&$2.14$&$1.57$&$1.86$&$0.21$&\\
		&$500$&$1$&$-$&$-$&$-$&$-$&$-$&$-$&$0.53$&\\
		&&$2$&$-0.50$&$1.52$&$2.03$&$1.72$&$1.39$&$1.25$&$0.25$&\\ 
		&&$3$&$-1.44$&$0.62$&$2.79$&$1.45$&$1.30$&$0.99$&$0.22$&\\ 
		&$1000$&$1$&$-$&$-$&$-$&$-$&$-$&$-$&$0.53$&\\
		&&$2$&$0.17$&$2.08$&$1.17$&$0.70$&$0.81$&$1.11$&$0.27$&\\
		&&$3$&$-1.93$&$-0.07$&$3.09$&$1.00$&$1.05$&$1.21$&$0.20$&\\
		\hline
	\end{tabular}	
\end{table}
\bigskip
\begin{table}[H]
	\centering\caption{True and Recovered values of the regression coefficients for $n = 200, 500, 1000$}\vspace{0.4cm}
	\label{tab:2}
	\begin{tabular}{r@{\hspace{.2in}}r@{\hspace{.2in}}r@{\hspace{.1in}}r@{\hspace{.1in}}r@{\hspace{.1in}}r@{\hspace{.1in}}r@{\hspace{.1in}}r@{\hspace{.1in}}r@{\hspace{.1in}}r@{\hspace{.1in}}}
		\hline
		Values&$n$&$g$ &$\beta_0$& $\beta_1$& $\beta_2$& $\beta_3$& $\beta_4$& $\beta_5$&\\ 
		\hline
		True&&$2$&$0.00$&$0.88$&$0.28$&$0.96$&$0.09$&$0.33$&\\ 
		&&$3$&$0.00$&$0.77$&$0.53$&$0.98$&$0.07$&$0.37$&\\ 
		\hline
		Recovered&$200$&$2$&0.33&0.73&0.26&0.87&0.07&0.39&\\
		&&$3$&0.30&0.71&0.29&0.84&0.18& 0.41&\\
		&$500$&$2$&0.25&0.82&0.29&0.92&0.09&0.32&\\ 
		&&$3$&0.24&0.75&0.42&0.89&0.086&0.38&\\ 
		&$1000$&$2$&$0.02$&$0.92$&$0.27$&$0.86$&$0.09$&$0.39$&\\
		&&$3$&$0.51$&$0.81$&$0.49$&$0.86$&$0.04$&$0.41$&\\
		\hline
	\end{tabular}
\end{table} 
\bigskip
\begin{table}[H]
	\centering\caption{The values of AIC, BIC, ICL of ZIPCWM for different $n = 200,500, 1000$ with true component $G=3$. For $n = 200$, all the selection criteria perform poorly. However, the performance gets improved with increased sample size.}\vspace{0.4cm}
	\label{tab:5}
	\begin{tabular}{r@{\hspace{.1in}}r@{\hspace{.1in}}r@{\hspace{.1in}}r@{\hspace{.1in}}r@{\hspace{.1in}}r@{\hspace{.1in}}r@{\hspace{.1in}}r@{\hspace{.1in}}r@{\hspace{.1in}}r@{\hspace{.1in}}r@{\hspace{.1in}}}
		\hline
		n&G&AIC&BIC&ICL&AWE&AIC3&AICc&AICu&Caic&\\ 
		\hline
		$200$&$2$&$2192.91$&$2080.77$&$2080.76$&$1798.62$&$2158.91$&$2178.48$&$2140.01$&$2046.77$&\\
		&$3$&$2055.82$&$1831.49$&$1822.53$&$1267.24$&$1987.82$&$1984.18$&$1899.56$&$1763.53$&\\
		&$4$&$1945.12$&$1608.69$&$1598.41$&$762.26$&$1843.12$&$1728.50$&$1583.78$&$1506.69$&\\
		&$5$&$\vec{1771.85}$&$\vec{1323.28}$&$\vec{1315.37}$&$\vec{194.70}$&$\vec{1635.85}$&$\vec{1180.35}$&$\vec{949.32}$&$\vec{1187.28}$&\\
		$500$&$2$&$7557.49$&$7414.20$&$7413.55$&$7100.90$&$7523.49$&$7552.37$&$7516.09$&$7380.20$&\\
		&$3$&$\vec{5128.59}$&$4841.99$&$4826.93$&$4215.40$&$\vec{5060.59}$&$\vec{5106.82}$&$5032.57$&$4773.99$&\\
		&$4$&$5212.08$&$4782.19$&$4750.51$&$3842.30$&$5110.08$&$5159.16$&$5043.82$&$4680.19$&\\
		&$5$&$5270.85$&$\vec{4697.66}$&$\vec{4649.80}$&$\vec{3444.47}$&$5134.85$&$5168.19$&$\vec{5008.09}$&$\vec{4561.66}$&\\
		$1000$&$2$&$12074.24$&$11907.38$&$11907.37$&$11570.51$&$12040.24$&$12071.78$&$12036.15$&$11873.38$&\\
		&$3$&$\vec{10234.37}$&$\vec{9863.09}$&$\vec{9900.64}$&$9226.91$&$\vec{10166.37}$&$\vec{10224.29}$&$\vec{10152.79}$&$\vec{9832.64}$&\\
		&$4$&$10860.90$&$10360.31$&$10268.41$&$9349.72$&$10758.90$&$10837.48$&$10728.78$&$10258.31$&\\
		&$5$&$10907.28$&$10239.83$&$10127.11$&$\vec{8892.37}$&$10771.28$&$10864.10$&$10716.76$&$10103.83$&\\
		\hline
	\end{tabular}
\end{table}
\begin{figure}[H]
	\centering
	\begin{minipage}[b]{0.45\textwidth}
		\includegraphics[width = 3in, height=2in]{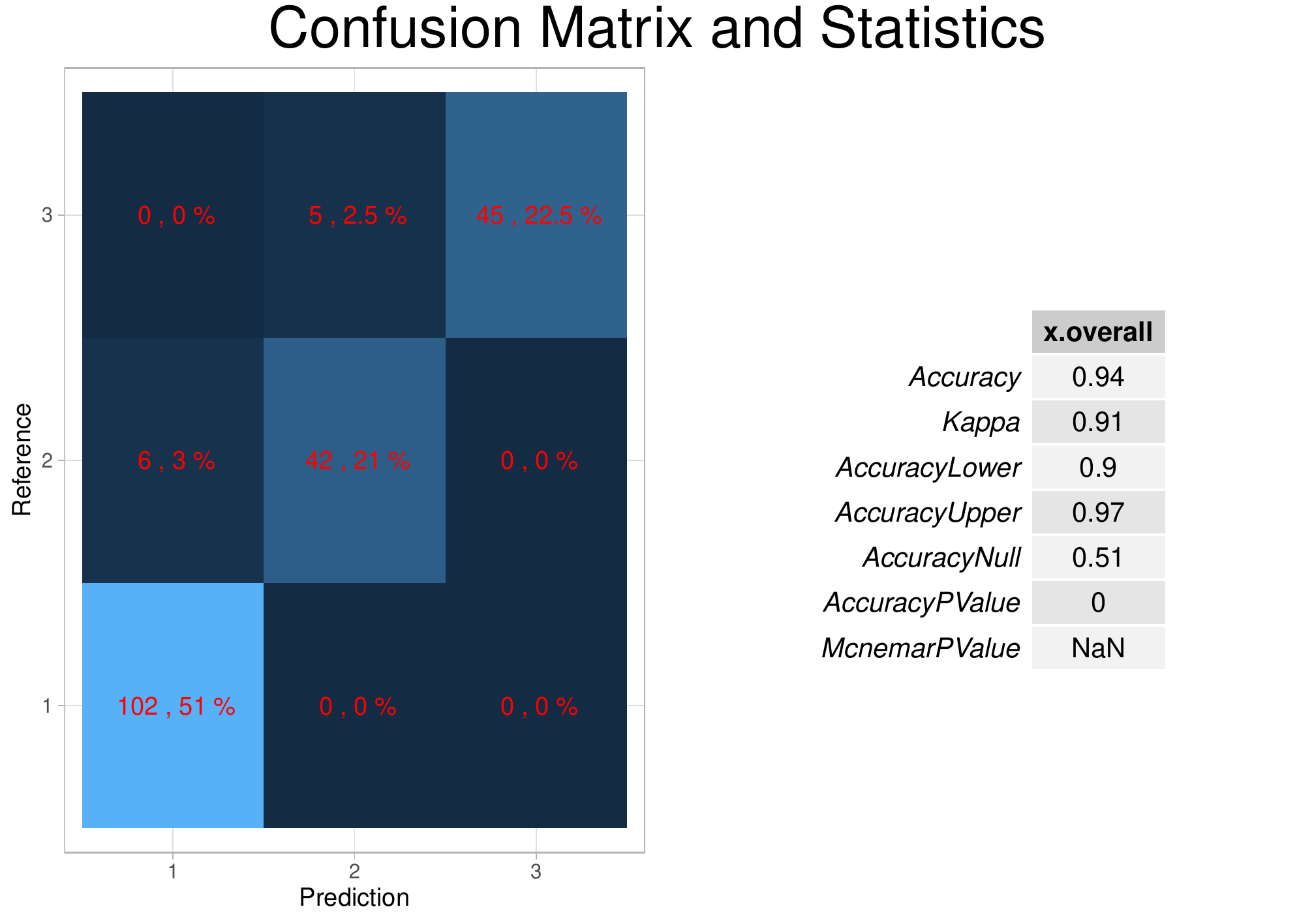}
		\centering\caption{\textit{The Visualization of Confusion Matrix and Statistics of the ZIPCWM with $n = 200$, $G = 3$.}} 
		\label{fig:1}
	\end{minipage}
	\hfil
	\begin{minipage}[b]{0.45\textwidth}
		\includegraphics[width = 3in, height=2in]{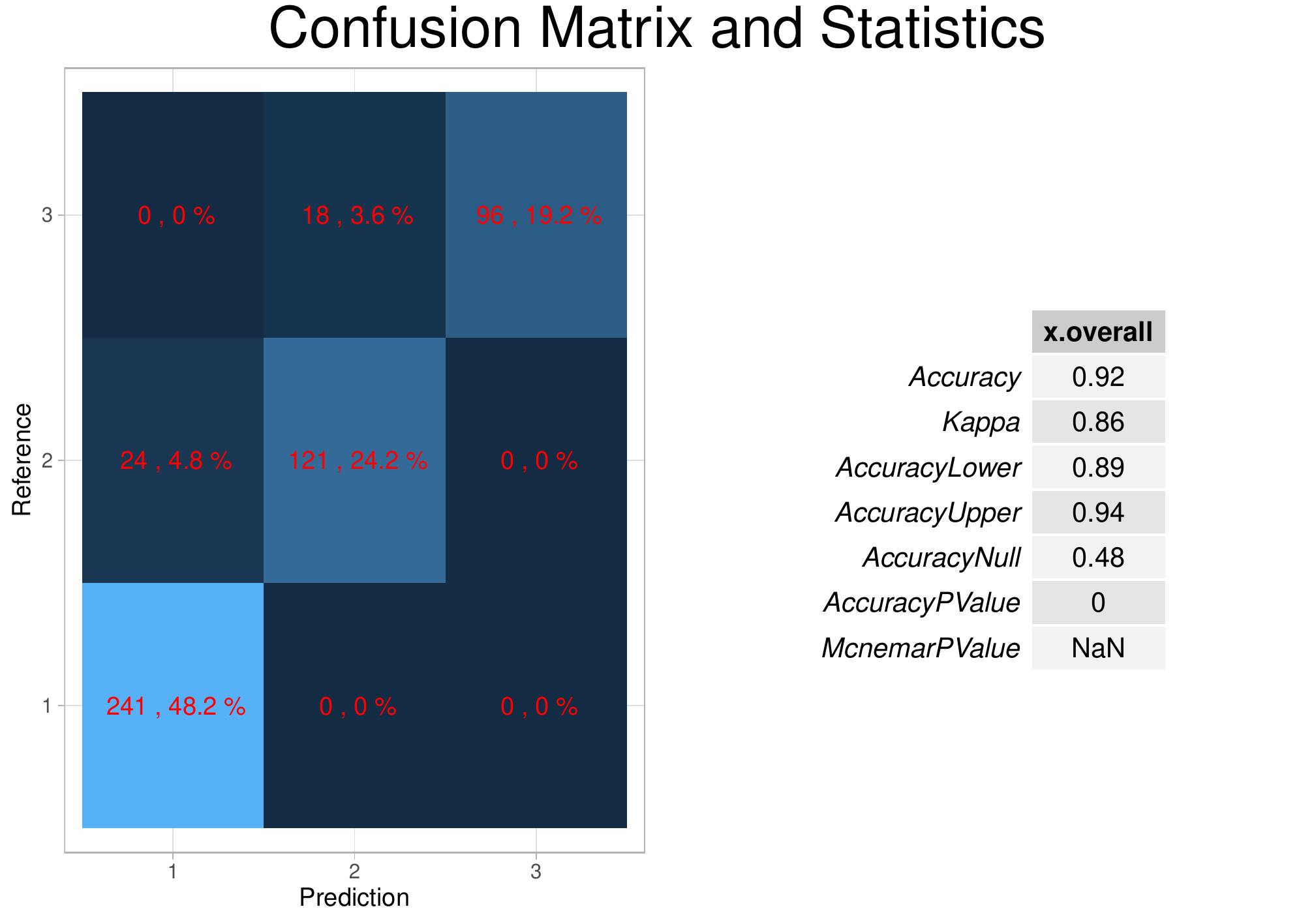}
		\caption{\textit{The Visualization of Confusion Matrix and Statistics of the ZIPCWM with $n = 500$, $G = 3$.}} 
		\label{fig:2}
	\end{minipage}
\end{figure}
\bigskip
\begin{figure}[H]
	\centering
	\begin{minipage}[b]{0.45\textwidth}
		\includegraphics[width = 3in, height=2in]{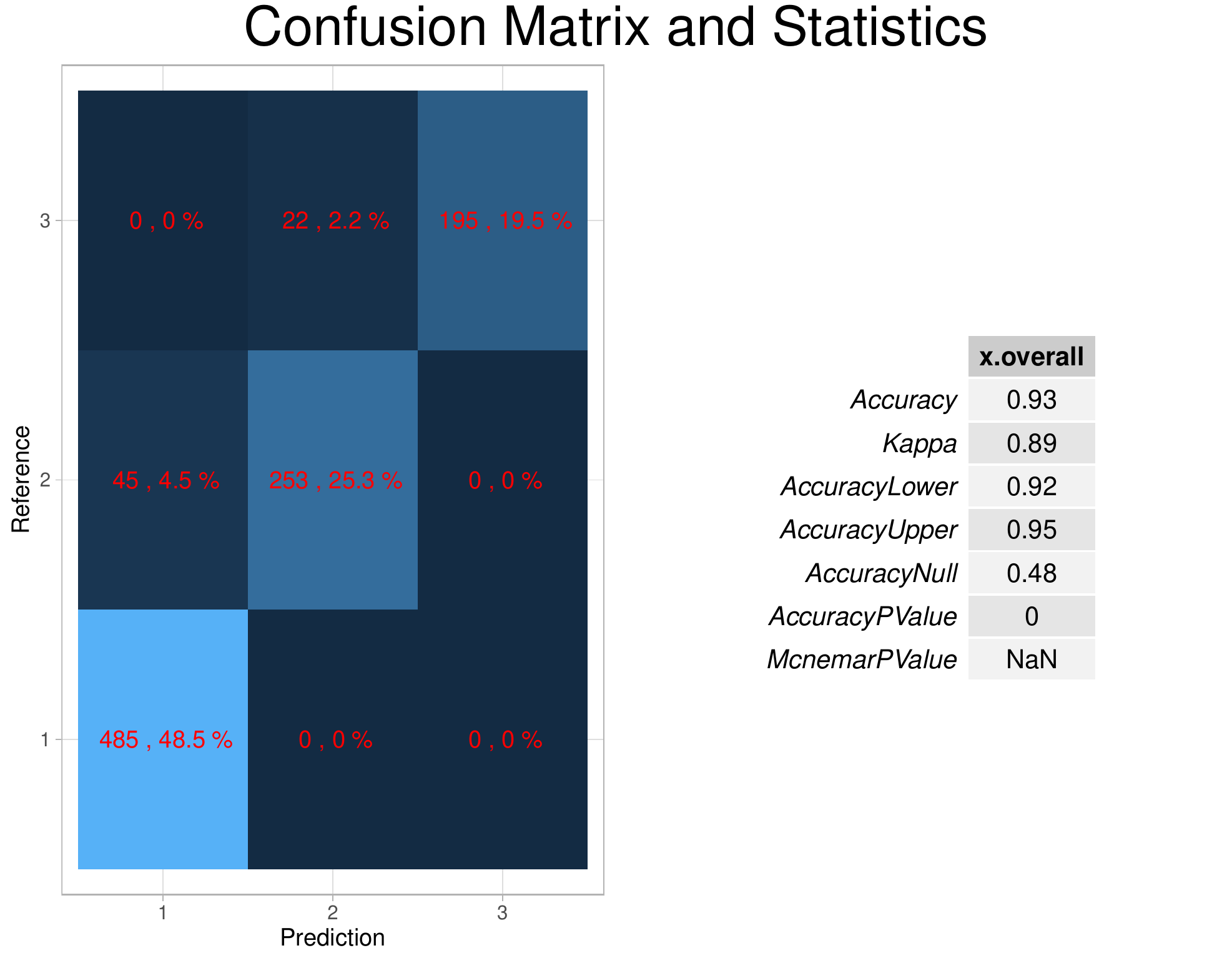}
		\centering\caption{\textit{The Visualization of Confusion Matrix and Statistics of the ZIPCWM with $n = 1000$, $G = 3$.}} 
		\label{fig:3}
	\end{minipage}
	\hfil
	\begin{minipage}[b]{0.45\textwidth}
		\includegraphics[width = 3in, height=2in]{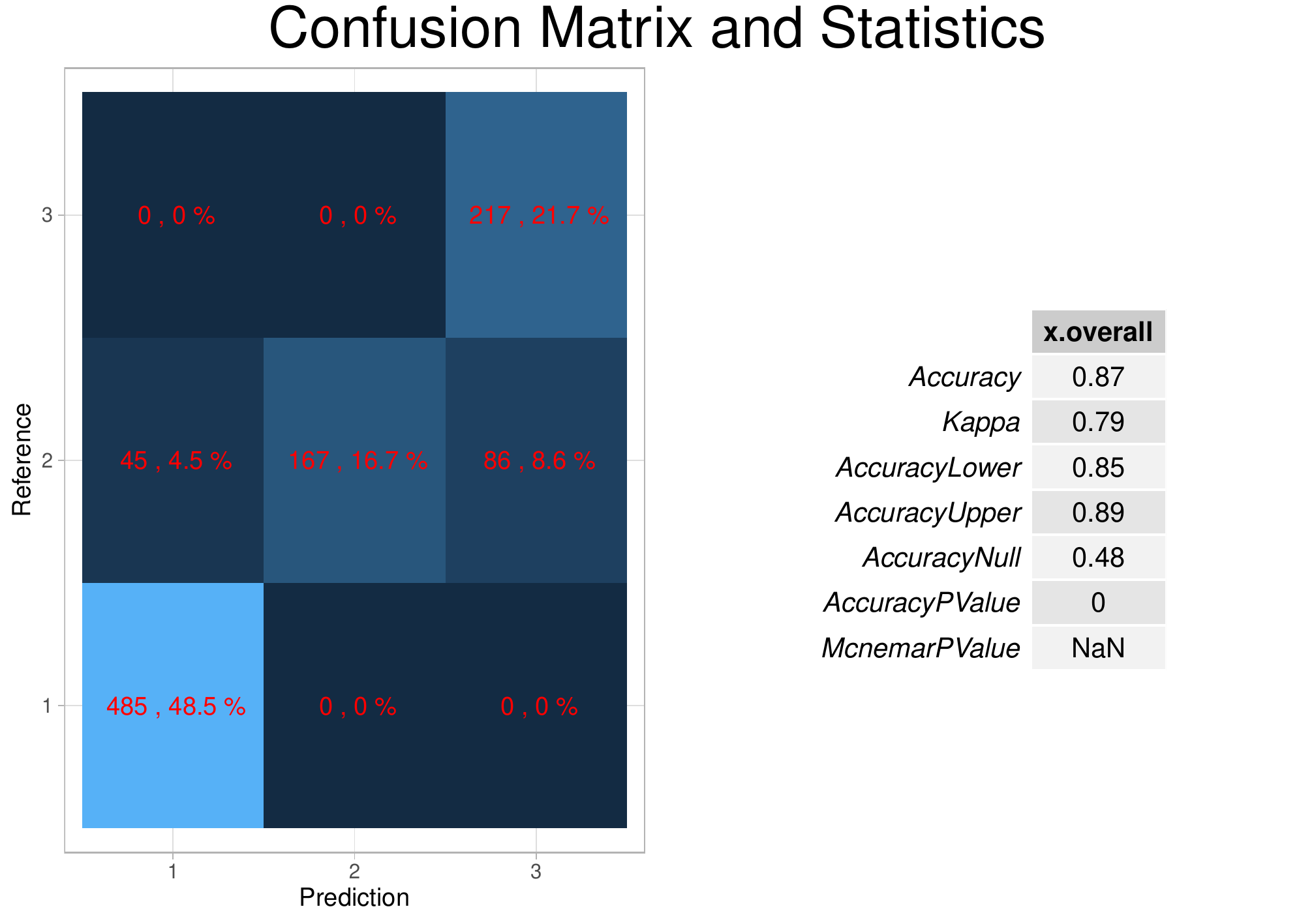}
		\caption{\textit{The Visualization of Confusion Matrix and Statistics of the FZIP regression mixture model with $n = 1000$, $G = 3$.}} 
		\label{fig:4}
	\end{minipage}
\end{figure}
\bigskip
\noindent components using eight different model selection criteria such as Akaike's information criterion (AIC) and Bayesian information criterion (BIC) model selection criteria, AWE, AICc, AIC3, AICu, Caic. The values presented are the average values of ten runs for each of the criterion. In Table \eqref{tab:5}, the effect of sample size is significantly evident. It can be seen that all the selection criteria select wrong number of components with small size. They all select the model with too many components (clusters) when the sample size is $n = 200$. However, when the sample size increases to $500$, AIC, AIC3, and AIC3 correctly select the true cluster component. Moreover, when the sample size is $100$, all the selection criteria except AWE pick the right choice of the model component. AWE displays a poor performance of overestimating the number of components in the data. All criteria except the AWE perform satisfactorily with higher sample size. Obviously, the performance of the criteria roughly gets improved with increasing sample size $n$. Table \eqref{tab:3} presents the misclassification rates by counting the number of observation classified into a different components to the original components. The overall misclassification rate is $5.5\%$ for $n = 200$, $8.4\%$ for $n = 500$, and $6.7\%$ for $n = 1000$, We also compare the result with the fixed zero-inflated Poisson mixture model with the misclassification rate of $13.1\%$. The classification power of fixed zero-inflated Poisson mixture model agrees with the simulated study provided in \cite{hwa2014}. PCWM achieves a slight higher classification accuracy of $89.8\%$ compared yo FZIP and misclassification rate of $10.2\%$. To validate the classification power, we report the ARI value for each model, which measures the agreement between the true cluster and classification result of the proposed model. The ARI of the ZIPCWM is $0.892$ when $n = 1000$ and FZIP has ARI of $0.847$. This means the predicted result of ZIPCWM agrees more with the true classification than the FZIP. The AUC for both sample sizes are $0.923$, while the AUC of FZIP model is $0.829$.
\begin{table}[H]
	\centering\caption{Confusion Matrix in the three component Model for $n = 200, 500, 1000$. For $1000$, ZIPCWM, PCWM, and FZIP are compared together. It was observed that ZIPCWM outperforms PCWM and FZIP.}\vspace{0.4cm}
	\label{tab:3}
	\begin{tabular}{r@{\hspace*{.1in}}r@{\hspace*{.1in}}r@{\hspace*{.1in}}r@{\hspace*{.1in}}r@{\hspace*{.1in}}rrr@{\hspace*{.1in}}r@{\hspace*{.1in}}r@{\hspace*{.1in}}r@{\hspace*{.1in}}r@{\hspace*{.1in}}rr}
		\hline
		$G$&Real Component&&&$1$&&$2$&&$3$&&& Misclass. rate (\%)&&\\ 
		\hline
		$200$&$1$&&&$102$&&$0$&&$0$&&&$0.00$&\\ 
		&$2$&&&$6$&&$42$&&$0$&&&$12.50$&\\
		&$3$&&&$0$&&$5$&&$45$&&&$10.00$&\\
		\hline
		Misclassification&&&$5.50$&\\
		Accuracy&&&$94.50\%$&\\
		\hline
		$500$&$1$&&&$241$&&$0$&&$0$&&&$0.00$&\\ 
		&$2$&&&$24$&&$121$&&$0$&&&$16.55$&\\
		&$3$&&&$0$&&$18$&&$96$&&&$15.79$&\\
		\hline
		rate&&&$8.40$&\\
		Accuracy&&&$91.60\%$&\\
		\hline
		$1000$&$1$&&&$485$&&$0$&&$0$&&&$0.00$&\\ 
		&$2$&&&$45$&&$253$&&$0$&&&$15.10$&\\
		&$3$&&&$0$&&$22$&&$195$&&&$10.14$&\\
		\hline
		Misclassification&&&$6.70$&\\
		Accuracy&&&$93.30\%$&\\
		\hline
		PCWM&$1$&&&$485$&&$0$&&$0$&&&$0.00$&\\ 
		&$2$&&&$15$&&$196$&&$87$&&&$34.23$&\\
		&$3$&&&$0$&&$0$&&$217$&&&$0.00$&\\
		\hline
		Overall &&&$10.20$&\\
		ccuracy&&&$89.80\%$&\\
		\hline
		FZIP&$1$&&&$485$&&$0$&&$0$&&&$0.00$&\\ 
		&$2$&&&$45$&&$167$&&$86$&&&$43.96$&\\
		&$3$&&&$0$&&$0$&&$217$&&&$0.00$&\\
		\hline
		Overall &&&$13.10$&\\
		ccuracy&&&$86.90\%$&\\
		\hline
	\end{tabular}
\end{table}
\bigskip
\section{Modeling of ZIPCWM on Real Data}\label{sec:6}
\subsection{The Use of Contraceptive Among married women}
This dataset is a subset of the $1987$ National Indonesia Contraceptive Prevalence Survey. The samples are married women who were either not pregnant or do not know if they were at the time of interview \cite{Dua2017}. The problem is to cluster the current contraceptive method choices (no use, long-term methods, or short-term methods) of a woman based on her demographic and socio-economic characteristics.
\noindent The demographic and socio-economic attributes are as follows:
Wife's age (numerical), Wife's education (categorical) 1=low, 2, 3, 4=high, Husband's education (categorical) 1=low, 2, 3, 4=high, Number of children ever born (numerical), Wife's religion (binary) 0=Non-Islam, 1=Islam, Wife's now working? (binary) 0=Yes, 1=No, Husband's occupation (categorical) 1, 2, 3, 4, Standard-of-living index (categorical) 1=low, 2, 3, 4=high, Media exposure (binary) 0=Good, 1=Not good, Contraceptive method used (class attribute) 1=No-use, 2=Long-term, 3=Short-term. The values of the method used range between $0$ and $3$. Approximately, $43\%$ of the sample values are equal to zero, which correspond to no use of contraceptive method.
\par First, we carried out a dispersion test to ascertain the absence of overdispersion in the data. The dispersion parameter is $0.83983$ which signifies the absence of overdispersion. Furthermore, the Poisson regression model was carried out to check the significant power of the variables in the data. The result shows that the wife's age has a significant effect on the use of contraception among married women. This is reasonable since the estimate $\beta_{\text{age}} < 0$ which is $-0.04375$. This means the younger married women are $0.09572$ times more expected to use contraception than their older married counterparts. The wife's education also contributes significantly to the use of contraception with a \textit{p-value} of $< 2e-16$. From the 
\begin{figure}[H]
	\centering
	\begin{minipage}[b]{0.45\textwidth}
		\includegraphics[width = 3in, height=2in]{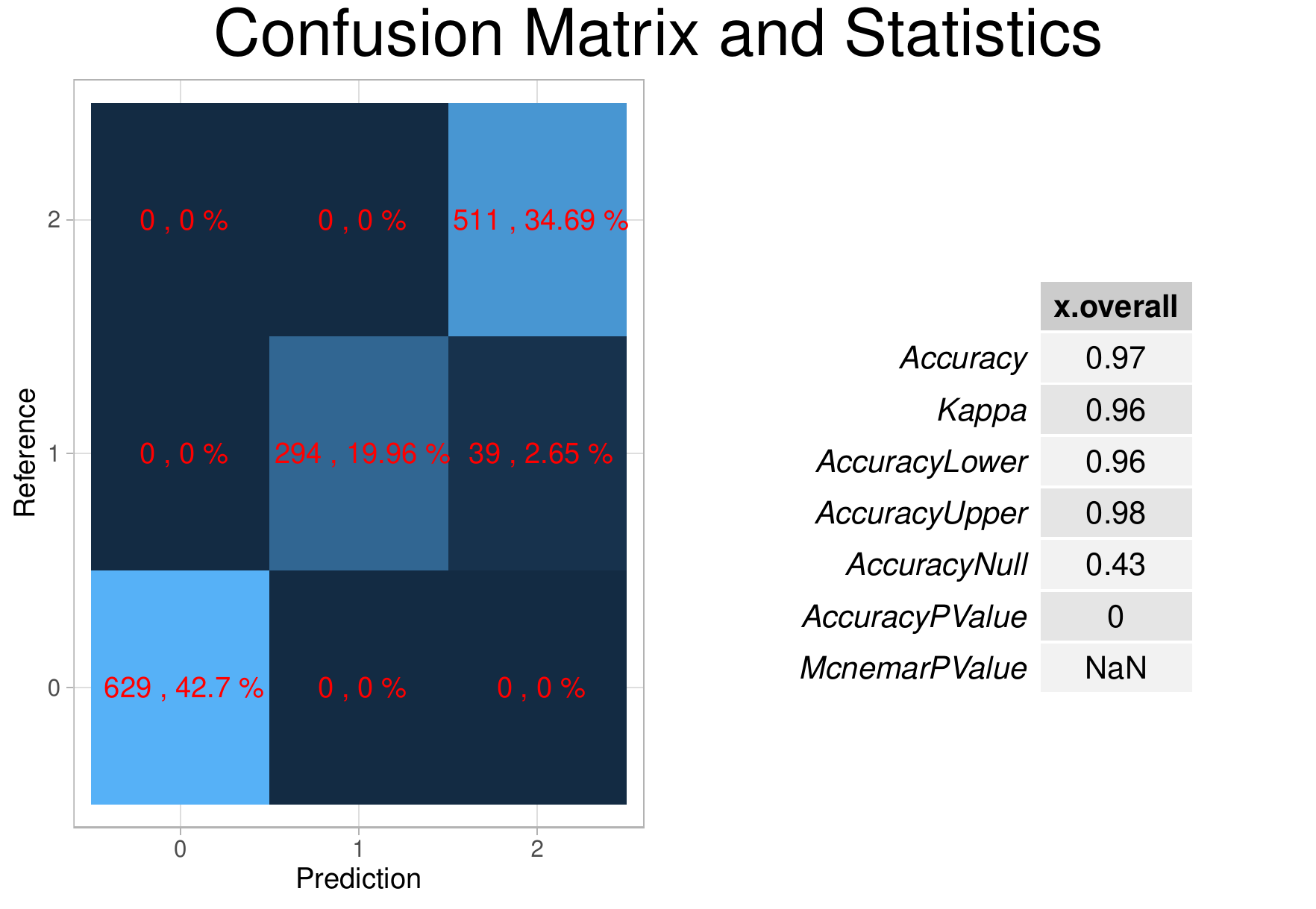}
		\centering\caption{\textit{The Visualization of Confusion Matrix and Statistics of the ZIPCWM that achieves $97\%$ classification accuracy}} 
		\label{fig:5}
	\end{minipage}
	\hfil
	\begin{minipage}[b]{0.45\textwidth}
		\includegraphics[width = 3in, height=2in]{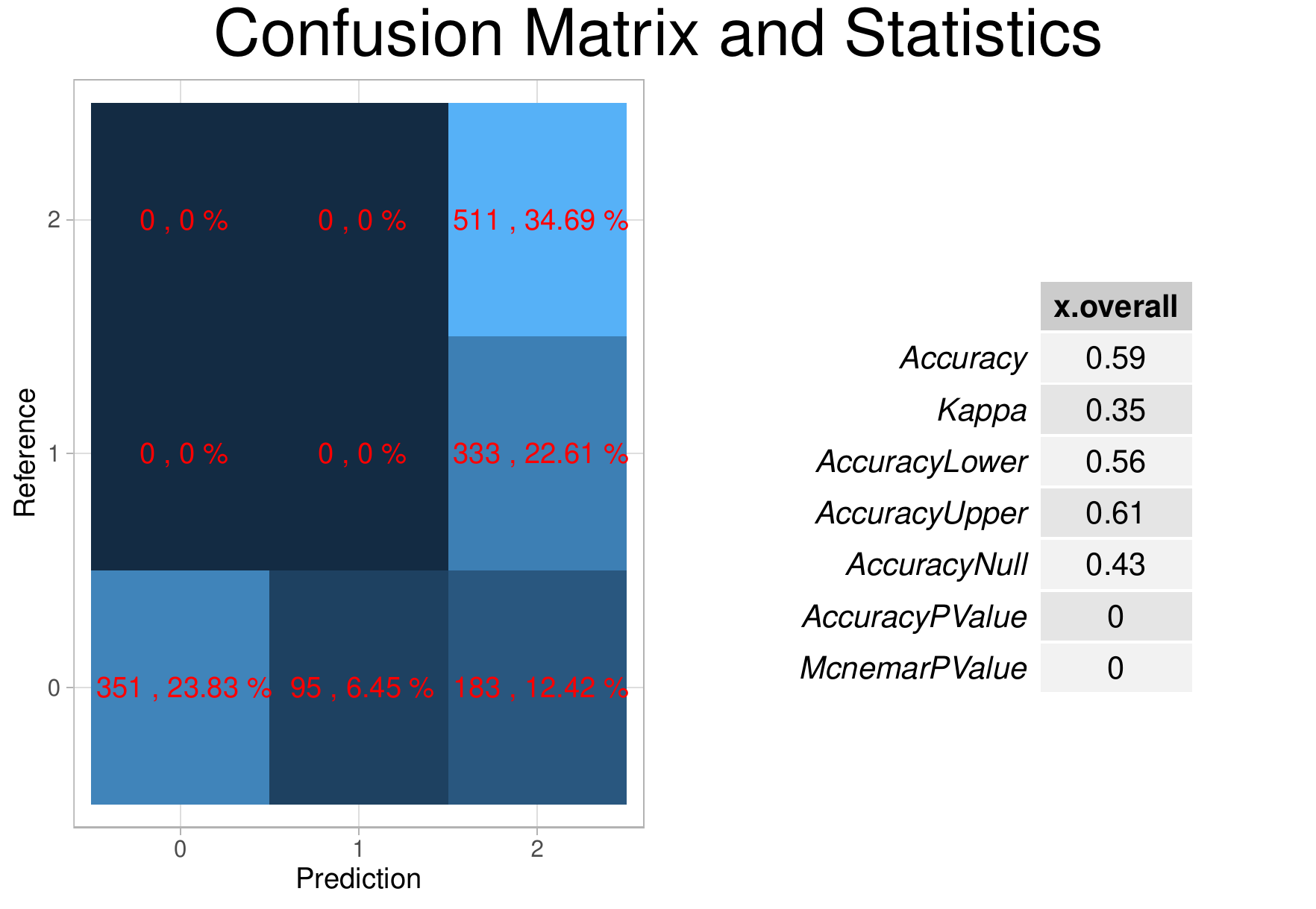}
		\caption{\textit{The Visualization of Confusion Matrix and Statistics of the PCWM that achieves about $59\%$ classification accuracy.}} 
		\label{fig:6}
	\end{minipage}
\end{figure}
\bigskip
\begin{table}[H]
	\centering\caption{Comparison of classification performance of ZIPCWM and PCWM with classification accuracy of $97\%$, and $58.5\%$ respectively. FZIP performs so poorly on the data as the model could not identify the reasonable component for the data.}\vspace{0.4cm}
	\label{tab:6}
	\begin{tabular}{r@{\hspace*{.1in}}r@{\hspace*{.1in}}r@{\hspace*{.1in}}r@{\hspace*{.1in}}r@{\hspace*{.1in}}rrr@{\hspace*{.1in}}r@{\hspace*{.1in}}r@{\hspace*{.1in}}r@{\hspace*{.1in}}r@{\hspace*{.1in}}r@{\hspace*{.1in}}}
		\hline
		Model&Real Component&&&$1$&&$2$&&$3$&&&Misclassification rate (\%)&\\ 
		\hline
		ZIPCWM&$1$&&&$629$&&$0$&&$0$&&&$0.00$&\\ 
		&$2$&&&$0$&&$294$&&$39$&&&$11.71$&\\
		&$3$&&&$0$&&$0$&&$511$&&&$0.00$&\\
		\hline
		Misclassification&&&$2.60$&&\\
		Accuracy&&&$97.40\%$&\\
		\hline
		PCWM&$1$&&&$351$&&$95$&&$183$&&&$44.20$&\\ 
		&$2$&&&$0$&&$0$&&$333$&&&$100.00$&\\
		&$3$&&&$0$&&$0$&&$511$&&&$0.00$&\\
		\hline
		Misclassification&&&$41.50$&\\
		Accuracy&&&$58.50\%$&\\
		\hline
	\end{tabular}
\end{table}
\bigskip
\noindent result, we deduce that well educated married women have $1.18$ more chance of using conception. The rest of the significant variables considered in the study are as follows; Number of children ever born, Standard-of-living index and Media exposure. We performed a comparison study of three different models such as Zero-Inflated Poisson Cluster-Weighted Models (ZIPCWM), Poisson Cluster-weighted Models (PCWM, \cite{Ingrassiaetal2015}), and Fixed Zero-Inflated Poisson Mixture models (FZIP, \cite{hwa2014}).
\par \noindent Here, the ZIPCWM, PCWM, and FZIP models with three components are fitted to the method of contraceptive among the married women based on their demographics and they are used to classify the data into three components. We computed the misclassification rate by counting the number of the incorrectly classified observation. Visualization results are presented in Figure \eqref{fig:5} and Figure \eqref{fig:6}. In Table \eqref{tab:6}, ZIPCWM has the overall misclassification rate is $2.6\%$. Moreover, it can be seen that most of the misclassifications are in component two which has $11.71\%$. In group one the number of married women according to the prediction of the model is $629$. The component two according to ZIPCWM has $294$ married women correctly classified as the married women that have long-term use of contraceptive. $39$ observations of the women that have long-term use were misclassified into cluster three (short-term use). 
ZIPCWM correctly predicts that $511$ married women have a short-term use of contraceptive. We observe that PCWM has a high overall misclassification rate of $41.50\%$. Moreover, all the component two was totally misclassified as component three. This is a problem of label switching. FZIP provides the worst classification. The model could not identify the components at all but classified all the women as coming from component three.
\begin{table}[H]
	\centering\caption{The significant student related variables}\vspace{0.4cm}
	\label{tab:7}
	\begin{tabular}{llllll}
		\hline
		\bf{Attribute}&&&&\bf{Description}&\\ 
		\hline
		\bf{school}&&&&student's school (binary: \textit{Gabriel Pereira} or \textit{Mousinho da Silveira})&\\
		\bf{sex}&&&&student's sex (binary: female or male)&\\
		\bf{age}&&&&student's age (numeric: from 15 to 22)&\\
		\bf{Pstatus}&&&&parent's cohabitation status (binary: urban or rural)&\\
		\bf{Medu}&&&&mother's education (numeric: 0 to 4)&\\
		\bf{Fedu}&&&&father's education (numeric: 0 to 4)&\\
		\bf{Mjob}&&&&mother's job (nominal)&\\
		\bf{Fjob}&&&&father's job (nominal)&\\
		\bf{guardian}&&&&student's guardian (nominal: "mother", "father" or "other")&\\
		\bf{studytime}&&&&weekly study time (numeric)&\\
		\bf{failures}&&&&number of past class failures (numeric: $n$ if 1 $\leq n < 3$, else 4)&\\
		\bf{famsize}&&&&family size (binary: $\leq 3$ or $>3$)&\\
		\bf{schoolsup}&&&&extra educational support (binary: yes or no)&\\
		\bf{famsup}&&&&family educational support (binary: yes or no)&\\
		\bf{paidclass}&&&&extra paid classes(binary: yes or no)&\\
		\bf{higher}&&&&wants to take higher education? (binary: yes or no)&\\
		\bf{romantic}&&&&with a romantic relationship (binary: yes or no)&\\
		\bf{famrel}&&&&quality of family relationships (numeric: from 1 - very bad to 5 - excellent)&\\
		\bf{freetime}&&&&free time after school (numeric: from 1 - very low to 5 - very high)&\\
		\bf{Dalc}&&&&workday alcohol consumption (numeric: from 1 - very low to 5 - very high)&\\
		\bf{Walc}&&&&weekend alcohol consumption (numeric: from 1 - very low to 5 - very high)&\\
		\bf{health}&&&&current health status (numeric: from 1 - very bad to 5 - very good)&\\
		\bf{absences}&&&&number of school absences (numeric: from 0 to 93)&\\
		\hline
	\end{tabular}
\end{table}
\subsection{Modeling the Number of Absence of students}\label{sec:6a}
The data consists of the performance of students in Portugal. The secondary education consists of $3$ years of schooling, preceding 9 years of basic education followed by the higher education. Due to some factors, most of the students join the public and free education system. The data was collected primarily with questionnaires. The structure of the education system in Portugal is a 20-point grading scale where 0 is the lowest and 20 is the perfect grade. The data was collected during the 2005-2006 school year from two public schools by \cite{cortezandsilva2008}. They designed the latter with closed questions related to demographic, social/emotional (\cite{pritchardandwilson2003}) and school related questions. The goal of the study is to classify the students according to the number of absence in the class. First of, we identify the significant factors that contribute to the absence. We modeled the data with a Poisson regression to investigate the significant effect of the variables on the presence of the students in class. Table \eqref{tab:7} shows the significant variables that contributed to the absence/presence of the students in class. It is interesting to know how some factors can strongly contribute to or negatively affect the success of the students such as parent cohabitation status. The response variable is taken to be a Poisson count. We also note the presence of class imbalance in the data i.e. one class makes up about $63\%$ of the response variable which makes it suitable for Zero-inflated model. The response variable is a number of event occurring (absence) in the class. The number of absence ranges from $0$ to $93$ where $0$ represents no absence and other numbers represent the number of absence. The response variable can also be seen as a binary variable of presence and absence. However, we performed a model selection test using the combination of eight model selection criteria with ZIPCWM and the result is presented below. Table \eqref{tab:8} confirms our intuition about the response variable. We observe that all the model selection criteria agree with the selection of the model with two components. This shows us that there are two categories of students in the class viz; those students that are regular in class and those that are irregular whether short-term or long-term.
\begin{figure}[H]
	\centering
	\begin{minipage}[b]{0.45\textwidth}
		\includegraphics[width = 3in, height=2in]{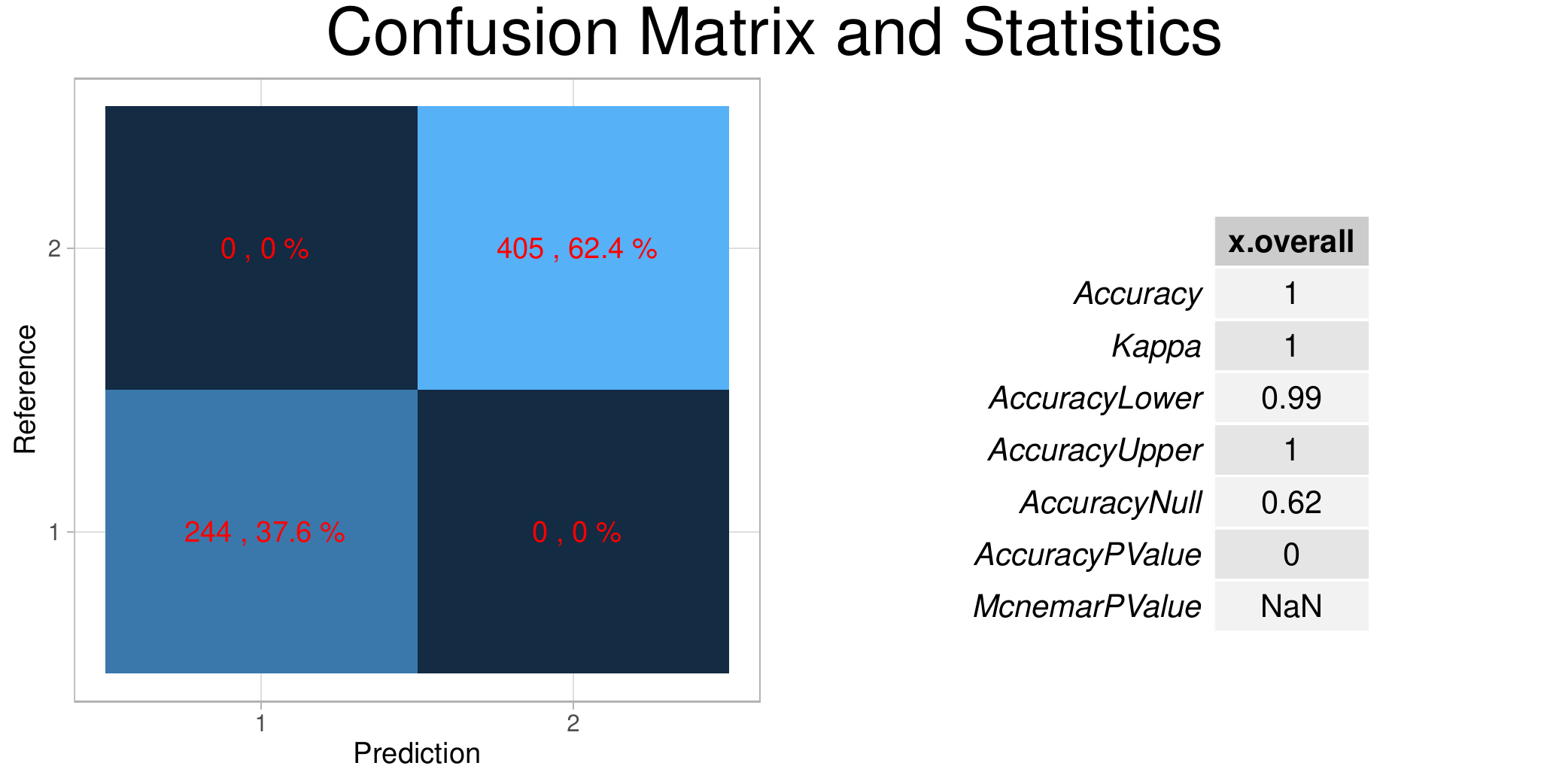}
		\centering\caption{\textit{The Visualization of Confusion Matrix and Statistics of the ZIPCWM that achieves $100\%$ classification accuracy.}} 
		\label{fig:9}
	\end{minipage}
	\hfil
	\begin{minipage}[b]{0.45\textwidth}
		\includegraphics[width = 3in, height=2in]{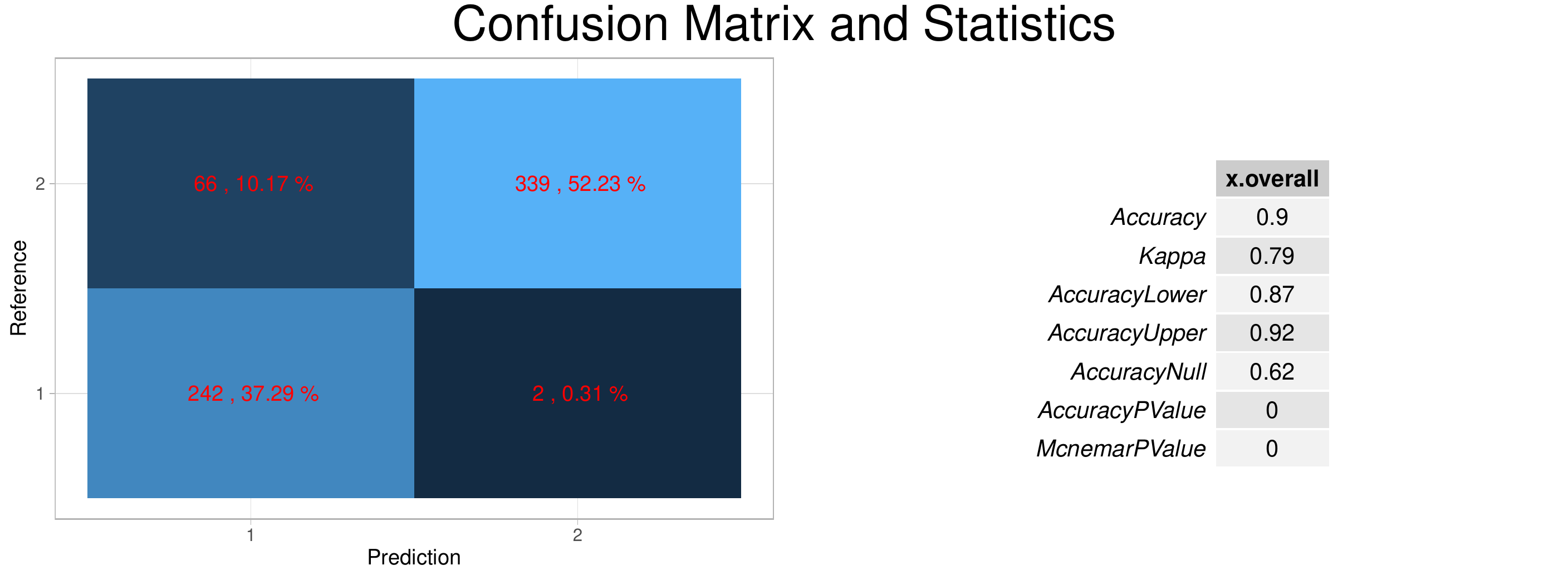}
		\caption{\textit{The Visualization of Confusion Matrix and Statistics of the PCWM that achieves about $90\%$ classification accuracy.}} 
		\label{fig:10}
	\end{minipage}
\end{figure}
\bigskip
\begin{table}[H]
	\centering\caption{The values of model selection criteria of ZIPCWM for different $G = 2,3, 4$ and $5$. Before proceeding with the clustering analysis, we carried out a model selection test. All the selection criteria used suggest the model with two components.}\vspace{0.4cm}
	\label{tab:8}
	\begin{tabular}{r@{\hspace{.1in}}r@{\hspace{.1in}}r@{\hspace{.1in}}r@{\hspace{.1in}}r@{\hspace{.1in}}r@{\hspace{.1in}}r@{\hspace{.1in}}r@{\hspace{.1in}}r@{\hspace{.1in}}r@{\hspace{.1in}}r@{\hspace{.1in}}}
		\hline
		&G&AIC&BIC&ICL&AWE&AIC3&AICc&AICu&Caic&\\ 
		\hline
		&$2$&$\vec{6505.81}$&$\vec{6420.77}$&$\vec{6419.84}$&$\vec{6240.74}$&$\vec{6486.81}$&$\vec{6504.60}$&$\vec{6484.28}$&$\vec{6401.77}$&\\
		&$3$&$7784.37$&$7614.30$&$7481.70$&$7254.24$&$7746.37$&$7779.51$&$7739.29$&$7576.30$&\\
		&$4$&$8059.74$&$7804.64$&$7633.21$&$7264.54$&$8002.74$&$8048.55$&$7987.79$&$7747.64$&\\
		&$5$&$7515.60$&$7175.47$&$7062.70$&$6455.34$&$7439.60$&$7495.14$&$7413.18$&$7099.49$&\\
		\hline
	\end{tabular}
\end{table}
\bigskip
\begin{table}[H]
	\centering\caption{Comparison of classification power of the ZIPCWM and PCWM Models. We intentionally omitted FZIP because we are interested in zero-inflated model against ordinary model. ZIPCWM provides higher classification accuracy relative to PCWM.}\vspace{0.4cm}
	\label{tab:9}
	\begin{tabular}{r@{\hspace*{.1in}}r@{\hspace*{.1in}}r@{\hspace*{.1in}}r@{\hspace*{.1in}}r@{\hspace*{.1in}}rrr@{\hspace*{.1in}}r@{\hspace*{.1in}}r@{\hspace*{.1in}}r@{\hspace*{.1in}}r@{\hspace*{.1in}}r@{\hspace*{.1in}}}
		\hline
		Model&Real Component&&&$1$&&&&$2$&&&Misclassification rate (\%)&\\ 
		\hline
		ZIPCWM&$1$&&&$244$&&&&$0$&&&$0.00$&\\
		&$2$&&&$0$&&&&$405$&&&$0.00$&\\
		\hline
		Misclassification&&&$0.00$&&\\
		Accuracy&&&$100.00\%$&&\\
		\hline
		PCWM&$1$&&&$242$&&&&$2$&&&$0.82$&\\ 
		&$2$&&&$66$&&&&$339$&&&$16.30$&\\
		\hline
		Misclassification&&&$10.48$&\\
		Accuracy&&&$89.52\%$&\\
		\hline
	\end{tabular}
\end{table}
\bigskip
\par \noindent After confirming the component of the model, we compare the classification power of ZIPCWM and PCWM models to distinguish between the zero-inflation model and ordinary model. We observe that the ZIPCWM outperforms PCWM. Table \eqref{tab:9} shows the misclassification rate of both models. ZIPCWM has the zero misclassification rate and achieves $100\%$ classification accuracy while the PCWM has $89.52\%$ classification accuracy. Also, Figure \eqref{fig:9} and Figure \eqref{fig:10} present the visualization of the confusion matrix with the classification accuracy. Thus, we conclude that ZIPCWM has a high classification power compared to PCWM. We measure the agreement between actual class and predicted class using Adjusted Rand Index (ARI). ZIPCWM has ARI of $1$ while PCWM has ARI of $0.81$. 
\section{Concluding Remarks}\label{sec:7}
\par In this paper, we have introduced the Zero-Inflated Poisson cluster-weighted model for class imbalance in data commonly known as censored information in medical data. Zero-Inflated Poisson cluster weighted model is a generalized form of the previously existing models such as Poisson cluster weighted model, Fixed Zero-Inflated Poisson mixture model. ZIPCWM allows modeling the data with mixed-type covariates which is the combination of the finite discrete and continuous variables. This provides an advantage over the limitations of GZIP and FZIP. Furthermore, we have extensively described an Expectation-Maximization algorithm for parameter estimation. To investigate the parameter recovery of the algorithm and the performance of various model selection criteria to select the number of mixture components, explicit simulation studies were carried out. Our simulation studies show that the proposed model work satisfactorily and the estimation technique performs well. The results show that the ZIPCWM with three components provides the best fit. These results lend support to the use of cluster weighted model and establish that ZIPCWM provides better fitting performance than the PCWM and FZIP when explaining zero-inflated heterogeneous data. Additionally, the proposed model was applied to the real data and a comparison study of the classification performance with its special models was investigated. It was observed that the proposed model outperformed its special models. In this work, we used eight different model selection criteria to select the number of mixing components, $G$. 

\bibliographystyle{apalike}
\bibliography{biblio}
\end{document}